\documentclass[10pt,reprint,amsmath,amssymb,aps,pra,showpacs,longbibliography,superscriptaddress]{revtex4-1}
\usepackage[latin9]{inputenc}
\usepackage{amsmath}
\usepackage{amssymb}
\usepackage{graphicx}
\usepackage{esint}

\makeatletter

\newcommand*\LyXThinSpace{\,\hspace{0pt}}

\usepackage{amsfonts}
\usepackage{graphics}

\makeatother

\begin{document}
\title{Two-dimensional coherent spectroscopy of trion-polaritons and exciton-polaritons
in atomically thin transition metal dichalcogenides}
\author{Hui Hu}
\affiliation{Centre for Quantum Technology Theory, Swinburne University of Technology,
Melbourne 3122, Australia}
\author{Jia Wang}
\affiliation{Centre for Quantum Technology Theory, Swinburne University of Technology,
Melbourne 3122, Australia}
\author{Riley Lalor}
\affiliation{Centre for Quantum Technology Theory, Swinburne University of Technology,
Melbourne 3122, Australia}
\author{Xia-Ji Liu}
\affiliation{Centre for Quantum Technology Theory, Swinburne University of Technology,
Melbourne 3122, Australia}
\date{\today}
\begin{abstract}
We present a microscopic many-body calculation of the nonlinear two-dimensional
coherent spectroscopy (2DCS) of trion-polaritons and exciton-polaritons
in charge-tunable transition-metal-dichalcogenides monolayers placed
in an optical microcavity. The charge tunability leads to an electron
gas with nonzero density that brings brightness to the trion - a polaron
quasiparticle formed by an exciton with a nonzero residue bounded
to the electron gas. As a result, a trion-polariton is created under
strong light-matter coupling, as observed in the recent experiment
by Sidler \textit{et al.} {[}Nat. Phys. \textbf{13}, 255 (2017){]}.
We analyze in detail the structure of trion-polaritons, by solving
an extended Chevy ansatz for the trion quasiparticle wave-function.
We confirm that the effective light-matter coupling for trion-polaritons
is determined by the residue of the trion quasiparticle. The solution
of the full many-body polaron states within Chevy ansatz enables us
to microscopically calculate the nonlinear 2DCS spectrum of both trion-polaritons
and exciton-polaritons. We predict the existence of three kinds of
off-diagonal cross-peaks in the 2DCS spectrum, as an indication of
the coherence among the different branches of trion-polaritons and
exciton-polaritons. Due to the sensitivity of 2DCS spectrum to quasiparticle
interactions, our work provides a good starting point to explore the
strong nonlinearity exhibited by trion-polaritons in some recent exciton-polariton
experiments.
\end{abstract}
\maketitle

\section{Introduction}

Exciton-polaritons in microcavities are hybrid light-matter quasiparticles,
formed due to strong coupling between excitons and tightly confined
optical modes \cite{Deng2010,Carusotto2013,Byrnes2014}. Owing to
the half-matter, half-light nature, they open a research frontier
of polaritonics to explore novel nonlinear quantum phenomena that
are impossible to observe in linear optical systems and are difficult
to reach in pure matter systems. This potential is further amplified
by the recent manipulation of atomically thin transition metal dichalcogenides
(TMD) \cite{Novoselov2005,Wang2018,Berkelbach2018}, such as MoS$_{2}$,
WS$_{2}$, MoSe$_{2}$, and WSe$_{2}$. In these two-dimensional materials,
robust bright excitons of electrons and holes with relatively large
effective masses and large exciton binding energy dominate the optical
response even at room temperature. As a result, TMD monolayers are
promising candidates for ultrafast polariton-based nonlinear optical
integrated devices, such as ultra-low threshold lasers, fast and low-power
switches, and all-optical integrated quantum gates. For this purpose,
strong polariton nonlinearity is typically required. However, so far
it remains a challenge to obtain strong exciton-exciton interaction
and polariton-polariton interaction \cite{Hu2020}. 

In this respect, the recent observation of trion-polaritons in charge-tunable
MoSe$_{2}$ monolayers by Sidler \textit{et al.} received considerable
interest \cite{Sidler2017}. At first glance, the existence of trion-polaritons
is a surprise, since a trion is a \emph{fermionic} three-particle
bound-state of one hole and two electrons and therefore in principle
it should not be able to couple with bosonic photon of light. But
now, we understand that trions in charge-tunable monolayers are actually
the quasiparticles of Fermi polarons \cite{Efimkin2017}, which are
excitons (as impurities) dressed by the whole Fermi sea of an electron
gas \cite{Massignan2014,Schmidt2018,Wang2022PRL,Wang2022PRA}. Except
in the true trion limit (with vanishing electron gas density), where
the three-particle bound-state is recovered, the trion is better viewed
as a dressed exciton with a nonzero \emph{residue} that characterizes
the free motion of the exciton \cite{Sidler2017,Efimkin2017}. As
a result, cavity mode can indeed couple to the trion and lead to the
formation of trion-polaritons. The real surprise of trion-polaritons
comes with the observation that there seems to be a large nonlinearity
in the optical response, as revealed by the pump-probe measurement
\cite{Tan2020}. The understanding of such a large nonlinearity has
been the focus of several theoretical analyses \cite{Rana2021,BastarracheaMagnani2021,Song2022arXiv}.
Further experimental investigations are definitely needed. In particular,
a nonlinear four-wave-mixing measurement, such as the two-dimensional
coherent spectroscopy (2DCS) \cite{Jonas2003,Li2006,Cho2008,Hao2016NanoLett,Muir2022}
would be ideally suitable to quantitatively characterize the large
nonlinearity of trion-polaritons.

The purpose of this work is two-fold. On the one hand, we wish to
clarify the nature of trion-polaritons by carefully examining the
full \emph{many-body} Fermi polaron wave-functions of either exciton-polariton
or trion-polariton, with the use of the Chevy ansatz that describes
the one-particle-hole excitations of the Fermi sea \cite{Chevy2006}.
The variational Chevy ansatz \cite{Chevy2006} (or equivalently the
many-body $T$-matrix theory \cite{Hu2022}) has been previously used
to determine the self-energy and the spectral function of trion-polaritons
\cite{Sidler2017,BastarracheaMagnani2021}. However, a detail analysis
of the many-body wave-functions is of lack. Here, our strategy is
to follow the recent theoretical study of the wave-functions of the
three-particle trion bound state \cite{Zhumagulov2022,Tempelaar2019},
where a single excess electron is approximately used to simulate the
whole Fermi sea through the $k$-space discretization. Our calculation
is free from such a $k$-space approximation. A trade-off, however,
is the ignorance of the internal degree of freedom of the exciton
wave-function. This ignorance is fully justified by the large exciton
binding energy ($\sim500$ meV), which is at least ten times larger
than the trion binding energy in TMD monolayers ($\sim30$ meV) \cite{Wang2018}.
The internal structure of excitons then should only bring negligible
effects on the low-energy properties of trion-polaritons.

On the other hand, the full many-body Fermi polaron wave-functions
obtained within the Chevy ansatz approximation allow us to microscopically
calculate the 2DCS spectrum of trion-polaritons, in addition to that
of exciton-polaritons. The microscopic determination of the 2DCS spectrum
of an interacting many-body system is highly non-trivial \cite{Lindoy2022,Wang2022arXiv1,Wang2022arXiv2,Hu2022arXiv}.
Therefore, we would like to restrict ourselves to the case of a \emph{single}
trion-polariton or exciton-polariton in the system \cite{Hu2022arXiv}.
This rules out the possibility of addressing the interaction effect
between two trion-polaritons that is of major interest. However, our
calculation would capture the basic features of the 2DCS spectrum,
which could then be used to discriminate the possible interaction
effects between two trion-polaritons in future 2DCS measurements.

The rest of the paper is organized as follows. In the next section
(Sec. II), we outline the model Hamiltonian for the Fermi-polaron-polaritons
in TMD monolayers and present the many-body solutions by using the
Chevy ansatz approximation. In Sec. III, we discuss the structures
of trion-polaritons and exciton-polaritons and the optical responses
of both photons and excitons. In Sec. IV, we predict the the 2DCS
spectroscopy and discuss in detail the off-diagonal cross-peaks, which
show the coherence between exciton-polaritons and trion-polaritons.
Finally, Sec. V is devoted to conclusions and outlooks.

\section{Model Hamiltonian and the Chevy ansatz solution}

In charge-tunable TMD monolayers, tightly bound excitons formed by
electrons and holes near the $K$ (or $K'$) valley move in the Fermi
sea of an electron gas in other valley with a nonzero electron density
that corresponds to a Fermi energy at about $\varepsilon_{F}\sim10$
meV. Electrons in the electron gas have opposite spin with respect
to the electron inside excitons. Therefore, their effective interaction
with excitons is attractive and is well characterized by a contact
interaction with strength $U<0$, whose magnitude is tuned to yield
the trion binding energy $E_{T}\sim30$ meV \cite{Tempelaar2019}.
The TMD monolayers can be placed in the antinode of a planar photonic
microcavity, with cavity photon mode being tuned near resonance with
the excitonic and trionic optical transitions. 

\subsection{Model Hamiltonian}

We denote the cavity photon mode and the exciton by the creation (or
annihilation) field operators $a_{\mathbf{k}}^{\dagger}$ ($a_{\mathbf{k}}$)
and $X_{\mathbf{k}}^{\dagger}$ ($X_{\mathbf{k}}$), respectively.
The electrons in the electron gas are described by the creation and
annihilation field operators $c_{\mathbf{k}}^{\dagger}$ and $c_{\mathbf{k}}$.
The polariton system under consideration therefore can be well described
by a Fermi polaron model Hamiltonian ($\hbar=1$) \cite{Sidler2017},
\begin{eqnarray}
\mathcal{H} & = & \mathcal{H}_{aX}^{(0)}+\sum_{\mathbf{k}}\epsilon_{\mathbf{k}}c_{\mathbf{k}}^{\dagger}c_{\mathbf{k}}+U\sum_{\mathbf{qkp}}X_{\mathbf{k}}^{\dagger}c_{\mathbf{q-k}}^{\dagger}c_{\mathbf{q}-\mathbf{p}}X_{\mathbf{p}},\label{eq:Hami}\\
\mathcal{H}_{aX}^{(0)} & = & \sum_{\mathbf{k}}\left[\omega_{\mathbf{k}}a_{\mathbf{k}}^{\dagger}a_{\mathbf{k}}+\epsilon_{\mathbf{k}}^{X}X_{\mathbf{k}}^{\dagger}X_{\mathbf{k}}+\frac{\Omega}{2}\left(a_{\mathbf{k}}^{\dagger}X_{\mathbf{k}}+h.c.\right)\right].\label{eq:aXHami}
\end{eqnarray}
Here, $\epsilon_{\mathbf{k}}=k^{2}/(2m_{e})$, $\omega_{\mathbf{k}}=k^{2}/(2m_{\textrm{ph}})+\delta$
and $\epsilon_{\mathbf{k}}^{X}=k^{2}/(2m_{X})$ are the single-particle
energy dispersion relation of electrons, cavity photons and excitons,
respectively, with electron mass $m_{e}$, photon mass $m_{\textrm{ph}}\sim10^{-5}m_{e}$
and exciton mass $m_{X}\simeq2m_{e}$ in 2D TMD materials \cite{Wang2018};
$\delta$ is the photon detuning measured in relative to the exciton
energy level; and finally, $\Omega$ is the light-matter coupling
(i.e., Rabi coupling). We will restrict ourselves to the case that
the maximum number of exciton-polaritons is one, i.e.,
\begin{equation}
\sum_{\mathbf{k}}(a_{\mathbf{k}}^{\dagger}a_{\mathbf{k}}+X_{\mathbf{k}}^{\dagger}X_{\mathbf{k}})\leq1,
\end{equation}
which realizes the Fermi polaron limit. In contrast, the density of
the electrons ($n=\sum_{\mathbf{k}}c_{\mathbf{k}}^{\dagger}c_{\mathbf{k}}$)
is tunable, by adjusting the Fermi energy $\varepsilon_{F}$ through
gate voltage in the experiments \cite{Sidler2017,Tan2020}.

In the absence of the electron gas, the strong light-matter coupling
leads to the well-defined two branches of exciton-polaritons: the
lower polariton and upper polariton \cite{Deng2010,Carusotto2013,Byrnes2014}.
With the electron gas, one may naively anticipate the effective interactions
between lower (upper) polarities and the electron gas, and hence the
formation of two separate lower and upper branches of Fermi polarons.
However, the correct physical picture turns out to be the formation
of attractive and repulsive Fermi polarons of dressed excitons in
the first place, and then the coupling of Fermi polarons to the light.
For this reason, the trion-polaritons is better viewed as Fermi-polaron-polaritons
\cite{Sidler2017}, where the treatment of a trion as an attractive
Fermi polaron is explicitly emphasized.

\subsection{The Chevy ansatz solution}

To solve the model Hamiltonian in the case of one exciton-polariton,
let us take the following Chevy ansatz,
\begin{eqnarray}
\left|P\right\rangle  & = & \left(\phi_{0}X_{0}^{\dagger}+\tilde{\phi}_{0}a_{0}^{\dagger}+\sum_{\mathbf{k}_{p}\mathbf{k}_{h}}\phi_{\mathbf{k}_{p}\mathbf{k}_{h}}X_{-\mathbf{k}_{p}+\mathbf{k}_{h}}^{\dagger}c_{\mathbf{k}_{p}}^{\dagger}c_{\mathbf{k}_{h}}\right.\nonumber \\
 &  & \left.+\sum_{\mathbf{k}_{p}\mathbf{k}_{h}}\tilde{\phi}_{\mathbf{k}_{p}\mathbf{k}_{h}}a_{-\mathbf{k}_{p}+\mathbf{k}_{h}}^{\dagger}c_{\mathbf{k}_{p}}^{\dagger}c_{\mathbf{k}_{h}}\right)\left|\textrm{FS}\right\rangle ,
\end{eqnarray}
for the Fermi-polaron-polariton states with zero total momentum $\mathbf{K=0}$.
Here, the Fermi sea at zero temperature $\left|\textrm{FS}\right\rangle $
is obtained by filling the single-particle energy level $\epsilon_{\mathbf{k}}$
with $N$ electrons, from the bottom of the energy band up to the
energy $\varepsilon_{F}$. The hole momentum $\mathbf{k}_{h}$ and
the particle momentum $\mathbf{k}_{p}$ satisfy the constraints $\epsilon_{\mathbf{k}_{h}}\leqslant\varepsilon_{F}$
and $\epsilon_{\mathbf{k}_{p}}>\varepsilon_{F}$, respectively. The
energy of the whole Fermi sea is denoted as $E_{\textrm{FS}}$.

The ansatz involves the free motions of excitons and photons with
the amplitudes $\phi_{0}$ and $\tilde{\phi}_{0}$, respectively.
It also describes the one-particle-hole excitations of the Fermi sea
due to the inter-particle interaction of excitons and electrons, with
the amplitude $\phi_{\mathbf{k}_{p}\mathbf{k}_{h}}$. Although there
is no direct interaction between photons and electrons, for completeness
we include the terms $a_{-\mathbf{k}_{p}+\mathbf{k}_{h}}^{\dagger}c_{\mathbf{k}_{p}}^{\dagger}c_{\mathbf{k}_{h}}\left|\textrm{FS}\right\rangle $
with the amplitude $\tilde{\phi}_{\mathbf{k}_{p}\mathbf{k}_{h}}$.
These terms actually do not contribute to the ansatz due to the negligible
photon mass, since the related energy would be extremely large (i.e.,
$\omega_{\mathbf{k}}$ becomes very significant for nonzero $\mathbf{k}\neq0$).

Unlike the previous works that only minimize the ground-state energy
of the Chevy ansatz for the variational parameters ($\phi_{0}$, $\tilde{\phi}_{0}$,
$\phi_{\mathbf{k}_{p}\mathbf{k}_{h}}$ and $\tilde{\phi}_{\mathbf{k}_{p}\mathbf{k}_{h}}$)
or the self-energy of polaritons \cite{Sidler2017,Tan2020}, here
we are interested in solving all the many-body Fermi-polaron-polariton
states, by using an alternative exact diagonalization approach. To
this aim, we put the system - consisting of $N$ electrons and a single
exciton-polariton - onto a two-dimensional square lattice with $L\times L$
sites. The electron density then takes the value 
\begin{equation}
n=\frac{N}{(La)^{2}},
\end{equation}
where $a$ is the lattice spacing and unless specified otherwise is
set to be unity ($a=1$). We consider that the exciton, photon, and
electrons hop on the lattice only to the nearest neighbor with strengths
$t_{a}$, $t_{X}$ and $t_{c}$, respectively. Their single-particle
energy dispersion relations are then given by ($\tilde{\omega}_{\mathbf{k}}=\omega_{\mathbf{k}}-\delta$),
\begin{eqnarray}
\tilde{\omega}_{\mathbf{k}} & = & -2t_{a}\left[\cos\left(k_{x}\right)+\cos\left(k_{y}\right)\right]+4t_{a}\simeq\frac{k_{x}^{2}+k_{y}^{2}}{2m_{\textrm{ph}}},\\
\epsilon_{\mathbf{k}}^{X} & = & -2t_{X}\left[\cos\left(k_{x}\right)+\cos\left(k_{y}\right)\right]+4t_{X}\simeq\frac{k_{x}^{2}+k_{y}^{2}}{2m_{X}},\\
\epsilon_{\mathbf{k}} & = & -2t_{c}\left[\cos\left(k_{x}\right)+\cos\left(k_{y}\right)\right]+4t_{c}\simeq\frac{k_{x}^{2}+k_{y}^{2}}{2m_{e}},
\end{eqnarray}
where $m_{\textrm{ph}}\equiv1/(2t_{a}a^{2})$, $m_{X}\equiv1/(2t_{X}a^{2})$
and $m_{e}\equiv1/(2t_{c}a^{2})$ in the dilute limit ($n\rightarrow0$)
that is of interest. In the same limit, we have the relation
\begin{equation}
\varepsilon_{F}\simeq4\pi nt_{c}=\frac{4\pi N}{L^{2}}t_{c}.
\end{equation}
 It is also easy to see the relations $t_{a}/t_{c}=m_{e}/m_{\textrm{ph}}\sim10^{5}$
and $t_{X}/t_{c}=m_{e}/m_{X}\simeq1/2$. We assume the periodic boundary
condition, so the momentum $\mathbf{k}$ on the lattice takes the
values, 
\begin{equation}
\left(k_{x},k_{y}\right)=\left(\frac{2\pi n_{x}}{L},\frac{2\pi n_{y}}{L}\right),
\end{equation}
with the integers $n_{x},n_{y}=-L/2+1,\cdots-1,0,1,\cdots L/2$.

On the square lattice, we may identify that the Hilbert space of the
model Hamiltonian involves four different types of expansion basis
states (at zero polaron momentum), 
\begin{eqnarray}
\left|1\right\rangle  & = & X_{0}^{\dagger}\left|\textrm{FS}\right\rangle ,\\
\left|2\right\rangle  & = & a_{0}^{\dagger}\left|\textrm{FS}\right\rangle ,\\
\left|3\right\rangle _{\mathbf{k}_{p}\mathbf{k}_{h}} & = & X_{-\mathbf{k}_{p}+\mathbf{k}_{h}}^{\dagger}c_{\mathbf{k}_{p}}^{\dagger}c_{\mathbf{k}_{h}}\left|\textrm{FS}\right\rangle ,\\
\left|4\right\rangle _{\mathbf{k}_{p}\mathbf{k}_{h}} & = & a_{-\mathbf{k}_{p}+\mathbf{k}_{h}}^{\dagger}c_{\mathbf{k}_{p}}^{\dagger}c_{\mathbf{k}_{h}}\left|\textrm{FS}\right\rangle ,
\end{eqnarray}
It is straightforward to see that the dimension of the Hilbert space
is $D=2+2N(L^{2}-N)$. By using the expansion basis states, the Fermi-polaron-polariton
model Hamiltonian then is casted into a $D\times D$ Hermitian matrix,
with the following matrix elements ($\mathcal{H}_{ji}=\mathcal{H}_{ij}^{*}$),
\begin{eqnarray}
\left\langle 1\right|\mathcal{H}\left|1\right\rangle  & = & E_{\textrm{FS}}+nU,\\
\left\langle 1\right|\mathcal{H}\left|2\right\rangle  & = & \frac{\Omega}{2},\\
\left\langle 1\right|\mathcal{H}\left|3\right\rangle _{\mathbf{k}'_{p}\mathbf{k}'_{h}} & = & \frac{U}{L^{2}},\\
\left\langle 1\right|\mathcal{H}\left|4\right\rangle _{\mathbf{k}'_{p}\mathbf{k}'_{h}} & = & 0,
\end{eqnarray}
and

\begin{eqnarray}
\left\langle 2\right|\mathcal{H}\left|2\right\rangle  & = & E_{\textrm{FS}}+\delta,\\
\left\langle 1\right|\mathcal{H}\left|3\right\rangle _{\mathbf{k}'_{p}\mathbf{k}'_{h}} & = & 0,\\
\left\langle 2\right|\mathcal{H}\left|4\right\rangle _{\mathbf{k}'_{p}\mathbf{k}'_{h}} & = & 0,
\end{eqnarray}
and
\begin{widetext}
\begin{eqnarray}
_{\mathbf{k}_{h}\mathbf{k}_{p}}\left\langle 3\right|\mathcal{H}\left|3\right\rangle _{\mathbf{k}'_{p}\mathbf{k}'_{h}} & = & \left[E_{\textrm{FS}}+\epsilon_{\mathbf{k}_{p}}-\epsilon_{\mathbf{k}_{h}}+\epsilon_{-\mathbf{k}_{p}+\mathbf{k}_{h}}^{X}+nU\right]\delta_{\mathbf{k}_{p}\mathbf{k}'_{p}}\delta_{\mathbf{k}_{h}\mathbf{k}'_{h}}+\frac{U}{L^{2}}\left(\delta_{\mathbf{k}_{h}\mathbf{k}'_{h}}-\delta_{\mathbf{k}_{p}\mathbf{k}'_{p}}\right),\\
_{\mathbf{k}_{h}\mathbf{k}_{p}}\left\langle 3\right|\mathcal{H}\left|4\right\rangle _{\mathbf{k}'_{p}\mathbf{k}'_{h}} & = & \frac{\Omega}{2}\delta_{\mathbf{k}_{p}\mathbf{k}'_{p}}\delta_{\mathbf{k}_{h}\mathbf{k}'_{h}},\\
_{\mathbf{k}_{h}\mathbf{k}_{p}}\left\langle 4\right|\mathcal{H}\left|4\right\rangle _{\mathbf{k}'_{p}\mathbf{k}'_{h}} & = & \left[E_{\textrm{FS}}+\epsilon_{\mathbf{k}_{p}}-\epsilon_{\mathbf{k}_{h}}+\omega_{-\mathbf{k}_{p}+\mathbf{k}_{h}}\right]\delta_{\mathbf{k}_{p}\mathbf{k}'_{p}}\delta_{\mathbf{k}_{h}\mathbf{k}'_{h}}.
\end{eqnarray}
\end{widetext}

\begin{figure*}
\centering{}\includegraphics[width=0.8\textwidth]{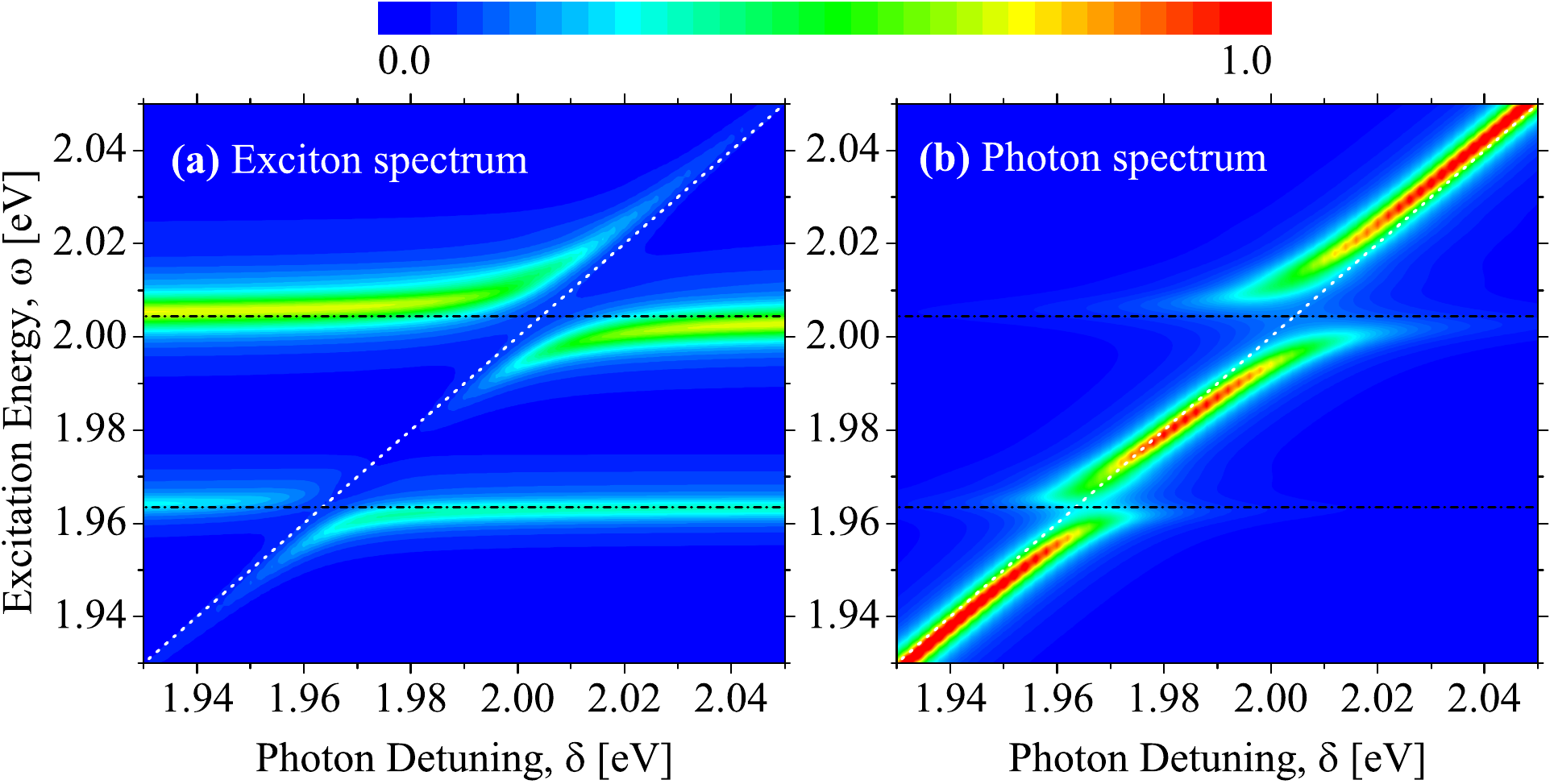}\caption{\label{fig:fig1_spectrum} Two-dimensional contour plots of the zero-momentum
spectral function of the exciton (a) and of the photon (b), as a function
of the photon detuning $\delta$ at the electron Fermi energy $\varepsilon_{F}=7.8$
meV. The two black horizontal dot-dashed lines show the energies of
the exciton (i.e., the repulsive polaron branch with $\varepsilon_{X}\simeq2004.4$
meV) and the trion (i.e., the attractive polaron branch with $\varepsilon_{T}\simeq1963.5$
meV), in the absence of the cavity photon field. The diagonal white
dotted line indicates the cavity photon detuning $\delta$. Two avoided
crossings at $\omega=\varepsilon_{X}$ and $\omega=\varepsilon_{T}$
are clearly visible. The spectral functions are measured in arbitrary
units and are plotted in a linear scale.}
\end{figure*}

We diagonalize the $D\times D$ Hermitian matrix to obtain all the
eigenvalues $E^{(n)}$ and eigenstates, from which we extract the
Fermi-polaron-polariton energies $\mathcal{E}^{(n)}=E^{(n)}-E_{\textrm{FS}}$,
the residue of excitons $Z_{X}^{(n)}\equiv\phi_{0}^{(n)*}\phi_{0}^{(n)}$
and the residue of photons $Z_{\textrm{ph}}^{(n)}\equiv\tilde{\phi}_{0}^{(n)*}\tilde{\phi}_{0}^{(n)}$.
Furthermore, we directly calculate the retarded Green functions of
excitons and photons,
\begin{eqnarray}
G_{X}\left(k=0,\omega\right) & = & \sum_{n}\frac{Z_{X}^{(n)}}{\omega-\mathcal{E}^{(n)}+\omega_{X}+i\delta},\\
G_{\textrm{ph}}\left(k=0,\omega\right) & = & \sum_{n}\frac{Z_{\textrm{ph}}^{(n)}}{\omega-\mathcal{E}^{(n)}+\omega_{X}+i\delta}.
\end{eqnarray}
and the associated spectral functions 
\begin{eqnarray}
A_{X}(k=0,\omega) & = & -\frac{1}{\pi}\textrm{Im}G_{X}(k=0,\omega),\\
A_{\textrm{ph}}(k=0,\omega) & = & -\frac{1}{\pi}\textrm{Im}G_{\textrm{ph}}(k=0,\omega).
\end{eqnarray}
Here, since we use a finite-size square lattice, the level spacing
in the single-particle dispersion relation is about $\delta=4t_{c}/L$.
We will use $\delta$ to replace the infinitesimal $0^{+}$ in the
spectral function and to eliminate the discreteness of the single-particle
energy levels. To make connection with the experimental measurement,
we measure the energy $\omega$ in the spectral function from the
top of the valence band by adding a constant energy shift $\omega_{X}=E_{g}-E_{X}=2$
eV \cite{Wang2018,Zhumagulov2022}, where $E_{g}$ and $E_{X}$ are
the band gap and the binding energy of excitons, respectively.

\section{Trion polaritons and the one-dimensional optical response}

In our numerical calculations, we consider a square lattice of $L=16$.
We set the hopping strength $t_{c}=10$ meV and then determine $t_{a}=t_{c}(m_{e}/m_{\textrm{ph}})=10^{6}$
meV and $t_{X}=t_{c}(m_{e}/m_{X})=5$ meV. At these parameters, the
spectral broadening factor $\delta=4t_{c}/L=2.5$ meV, which qualitatively
agrees the homogeneous broadening observed in the optical response
of exciton-polaritons \cite{Wang2018}. We take an attractive interaction
strength $U=-8t_{c}=-80$ meV, which leads to a trion energy at about
$-3.2t_{c}=-32$ meV in the dilute limit (i.e., $n\rightarrow0$ or
$N=1$ at $L=16$), in reasonable agreement with the trion binding
energy $E_{T}\sim30$ meV found in 2D TMD materials \cite{Wang2018}. 

Most of our calculations are carried out for a number of electrons
$N=16$, which corresponds to a Fermi energy $\varepsilon_{F}\simeq4\pi Nt_{c}/L^{2}\simeq7.8$
meV \cite{Sidler2017,Tan2020}. At this number of electrons, we find
the attractive polaron energy $E_{A}\simeq-3.65t_{c}=-36.5$ meV and
the repulsive polaron energy $E_{R}\simeq+0.44t_{c}=4.4$ meV, without
the cavity field. Measured from the top of the valence band, these
values give rise to the trion energy $\varepsilon_{T}=E_{A}+\omega_{X}=1963.5$
meV and the exciton energy $\varepsilon_{X}=E_{R}+\omega_{X}=2004.4$
meV. 

For convenience, we will also measure the photon detuning $\delta$
with respect to the top of the valence band, and understand it to
be $\delta+\omega_{X}$ without any confusion. For the light-matter
coupling, we always fix the Rabi frequency to be $\Omega=2t_{c}=20$
meV.

\begin{figure}
\begin{centering}
\includegraphics[width=0.45\textwidth]{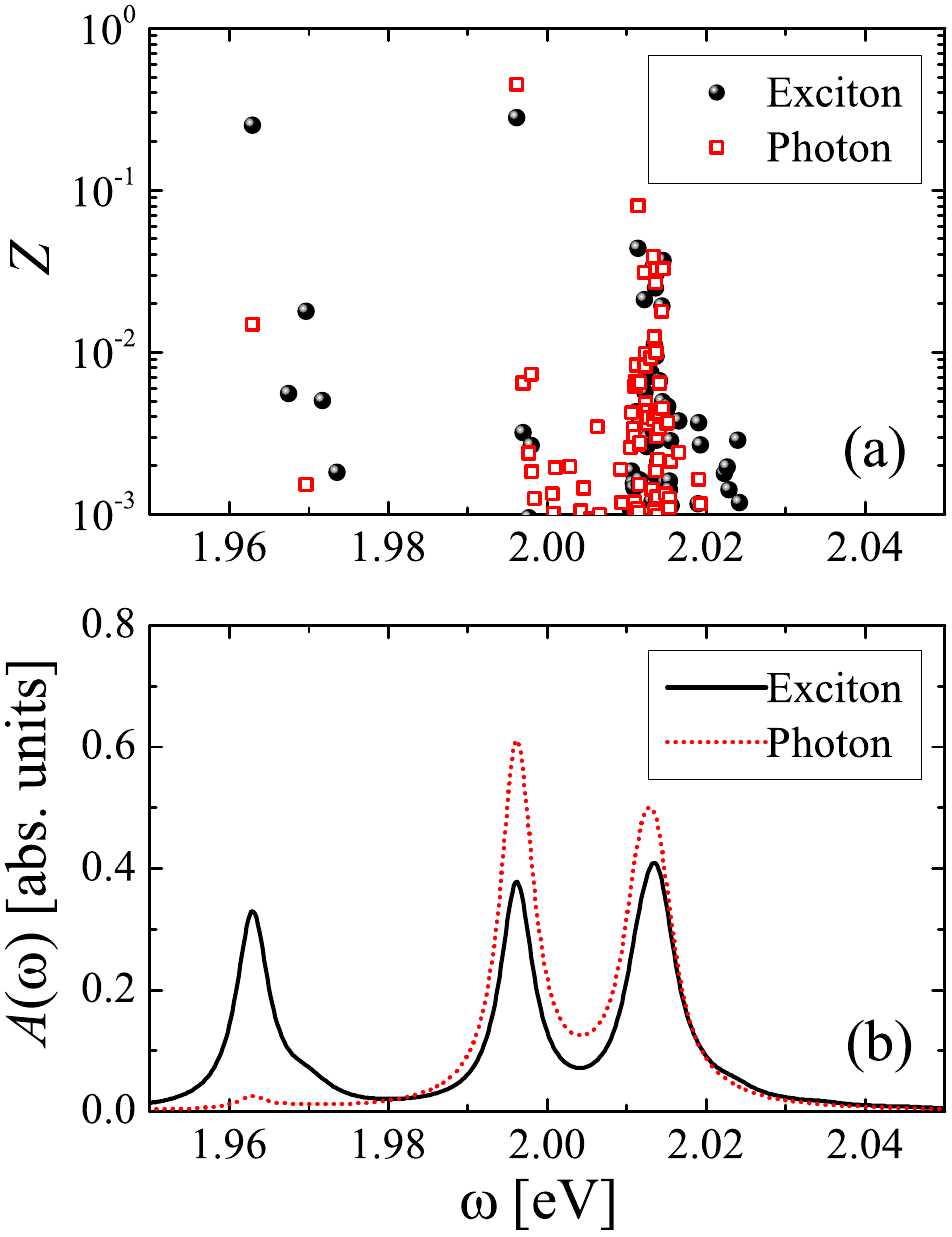}
\par\end{centering}
\centering{}\caption{\label{fig:fig2_ResidueExcitonCrossing} (a) Residues of the exciton
(black solid circles) and the photon (red empty square) for each many-body
state that is arranged with increasing energy. (b) The spectral function
of the exciton (black solid line) and the photon (red dotted line),
shown in arbitrary units. Here, we take a cavity photon detuning $\delta=\varepsilon_{X}\simeq2004.4$
meV. The electron Fermi energy is $\varepsilon_{F}=7.8$ meV.}
\end{figure}

\begin{figure}
\begin{centering}
\includegraphics[width=0.45\textwidth]{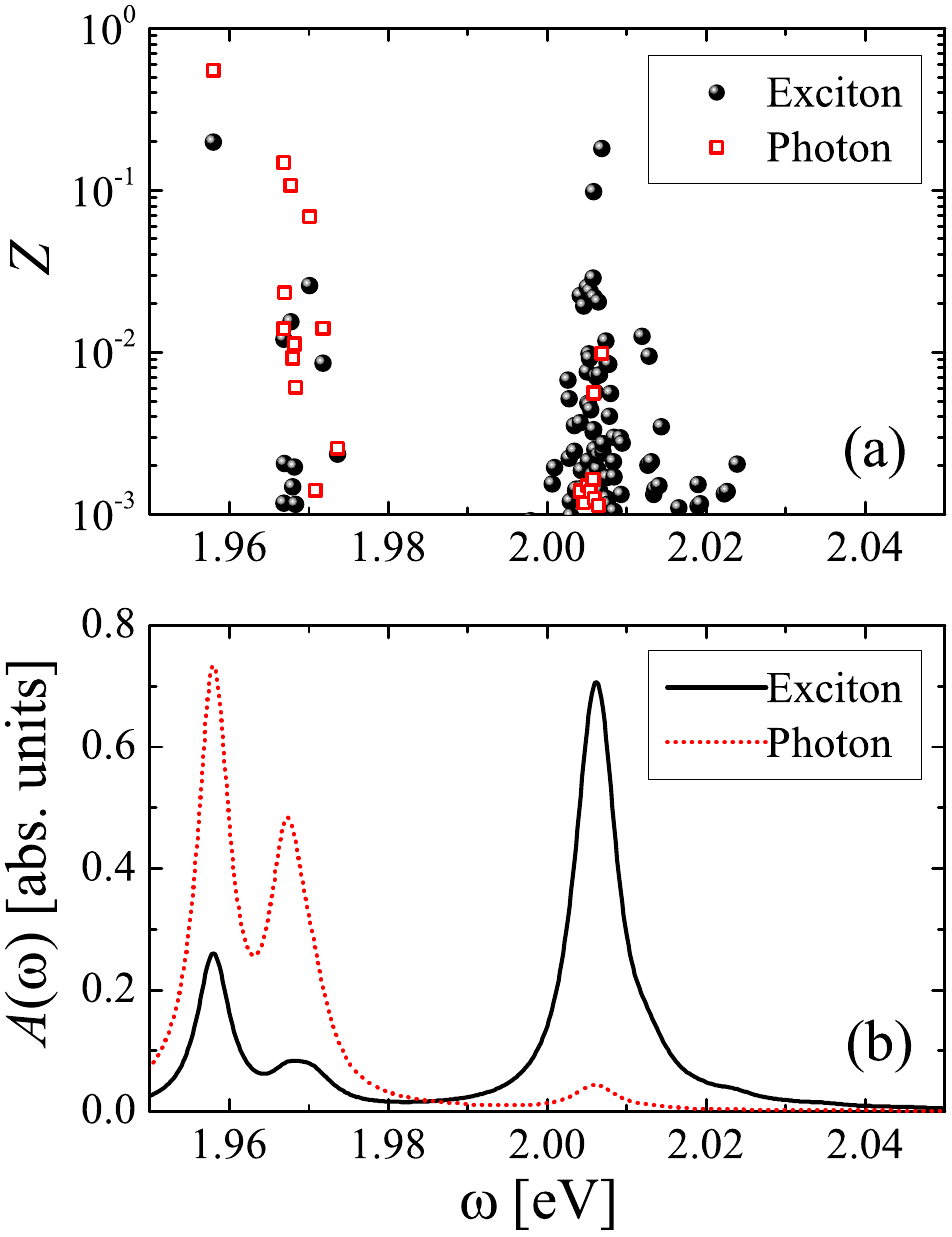}
\par\end{centering}
\centering{}\caption{\label{fig:fig3_ResidueTrionCrossing} (a) Residues of the exciton
(black solid circles) and the photon (red empty square) for each many-body
state that is arranged with increasing energy. (b) The spectral function
of the exciton (black solid line) and the photon (red dotted line),
shown in arbitrary units. Here, we take a photon detuning $\delta=\varepsilon_{T}\simeq1963.5$
meV. The electron Fermi energy is $\varepsilon_{F}=7.8$ meV.}
\end{figure}

In Figs. \ref{fig:fig1_spectrum}(a) and \ref{fig:fig1_spectrum}(b),
we report the zero-momentum spectral functions $A_{X}\left(k=0,\omega\right)$
and $A_{\textrm{ph}}\left(k=0,\omega\right)$ at the typical experimental
Fermi energy $\varepsilon_{F}=7.8$ meV for excitons and photons,
respectively, in the form of the two-dimensional contour plot with
a linear scale (as indicated on the top of the figure). Both spectral
functions clearly show an avoided crossing at the energy close to
$\omega_{X}=2$ eV. The two branches can be well-understood as the
upper and lower polaritons given by the model Hamiltonian $\mathcal{H}_{aX}^{(0)}$,
which exist even in the absence of the electron gas. This is evident
if we compare Fig. \ref{fig:fig1_spectrum} with Fig. \ref{fig:figS1_spectrum}
in Appendix A, where the latter figure reports the results at a much
smaller Fermi energy $\varepsilon_{F}=1$ meV. For the upper and lower
polariton branches, we find that the existence of the electron gas
will slightly shift the position of the avoided crossing (i.e., from
$\omega_{X}=2000$ meV to $\varepsilon_{X}\simeq2004.4$ meV), due
to the exciton-electron interaction that becomes effectively repulsive
for the excited state of repulsive polarons. 

The main effect of the electron gas to the spectral functions is the
appearance of an additional avoided crossing, the trion-polariton,
at the trion energy $\omega=\varepsilon_{T}$. At the Fermi energy
$\varepsilon_{F}=7.8$ meV in Fig. \ref{fig:fig1_spectrum}, this
avoided crossing has an energy splitting smaller than but comparable
to the Rabi coupling $\Omega=20$ meV for the exciton-polariton. The
shape of the avoided crossing is apparently asymmetric in the exciton
spectrum. At the much smaller Fermi energy $\varepsilon_{F}=1$ meV
in Fig. \ref{fig:figS1_spectrum}, the avoided crossing can hardly
be identified in both exciton spectrum and photon spectrum, which
unambiguously indicates that the existence of a Fermi sea is the key
source for the trion-polariton.

\begin{figure}
\begin{centering}
\includegraphics[width=0.45\textwidth]{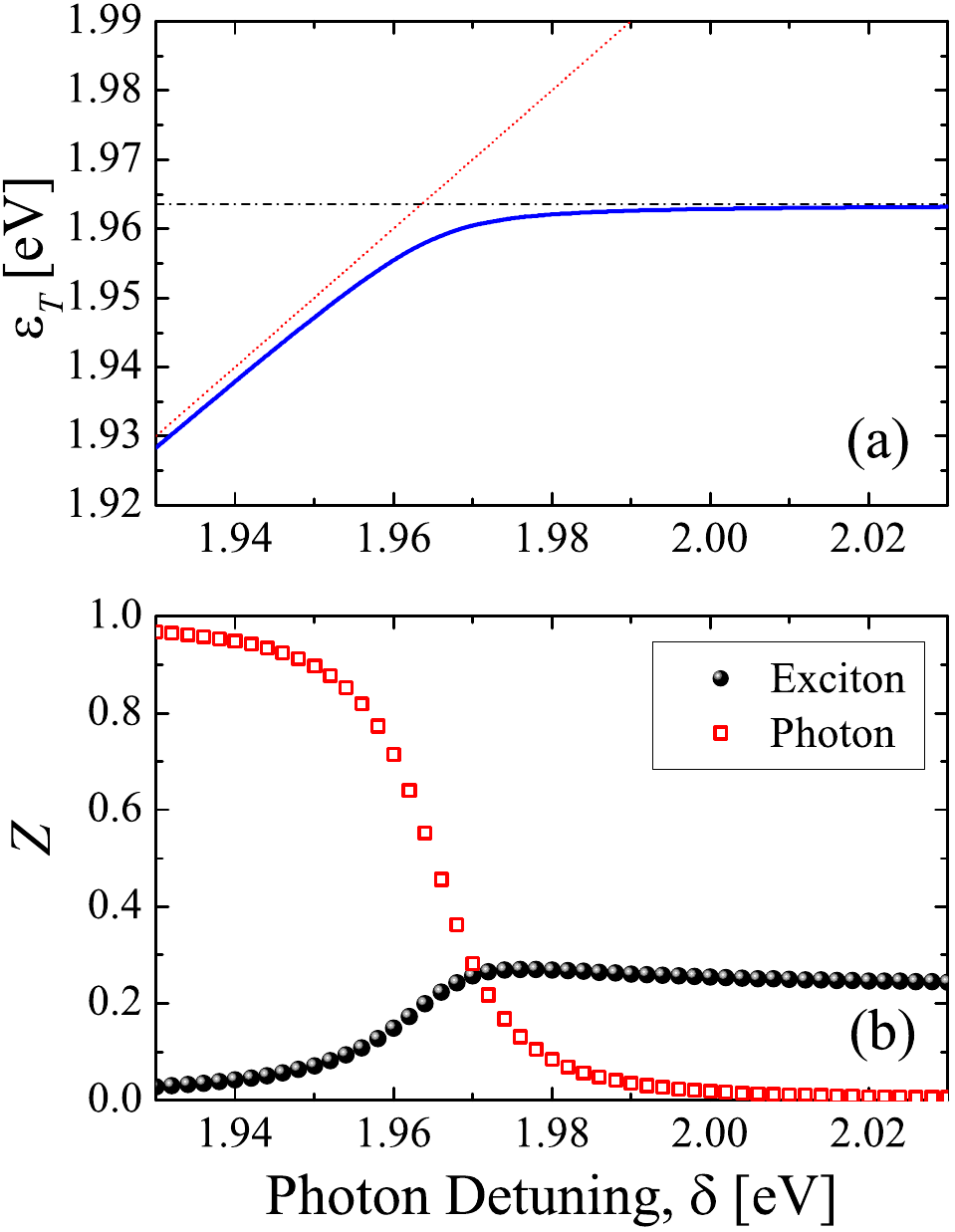}
\par\end{centering}
\centering{}\caption{\label{fig:fig4_GroundStatePolaron} (a) The ground-state energy of
Fermi-polaron-polaritons as a function of the photon detuning. The
red dotted line and the black dot-dashed line show the cavity photon
detuning and the trion energy without cavity field $\varepsilon_{T}\simeq1963.5$
meV. (b) Residues of the exciton (black solid circles) and the photon
(red empty squares) of the ground-state as a function of the photon
detuning. Here, we take the electron Fermi energy $\varepsilon_{F}=7.8$
meV.}
\end{figure}

To better understand the two avoided crossings for exciton-polaritons
and trion-polaritons, we show in Fig. \ref{fig:fig2_ResidueExcitonCrossing}
and Fig. \ref{fig:fig3_ResidueTrionCrossing} the residues (upper
panel) and spectral functions (lower panel) of excitons and photons,
at the photon detuning $\delta=\varepsilon_{X}$ and $\delta=\varepsilon_{T}$,
respectively.

Let us first focus on the avoided crossing for exciton-polaritons
at $\delta=\varepsilon_{X}\simeq2004.4$ meV in Fig. \ref{fig:fig2_ResidueExcitonCrossing}.
The composition of the different branches might be seen from the exciton
and photon residues. The upper branch (or the rightest branch) locates
at the energy $\sim2.013$ eV and consists of a number of many-body
energy levels that distribute nearby with notable exciton and photon
residues. For this upper branch, due to its \emph{collective} nature,
it seems difficult to find a Hopfield coefficient that clearly defines
the contributions or components from cavity photons and excitons,
as in the case of conventional exciton-polaritons. In contrast, for
the lower branch (or the middle branch in the range of the whole plot,
which is referred to as middle polariton in the literature) located
at the energy $\sim1.996$ eV, we find that it is only contributed
by one dominated state. All other nearby many-body states have residues
much less than $1\%$. This branch seems to decouple from the particle-hole
excitations of the Fermi sea and therefore retains the characteristic
of the exciton-polariton without the electron gas. We note that, the
energy splitting between the upper and lower branches is given by
$2.013-1.996=0.017$ eV or 17 meV, which is slightly smaller than
the Rabi coupling $\Omega=20$ meV. We attribute this slight difference
to the transfer of the residue or the oscillator strength to the third
branch (the lowest-energy branch) in the exciton spectrum, as shown
in Fig. \ref{fig:fig2_ResidueExcitonCrossing}(b).

The situation for the avoided crossing of trion-polaritons at $\delta=\varepsilon_{T}$
is very similar. As can be seen from Fig. \ref{fig:fig3_ResidueTrionCrossing},
the upper branch of this avoided crossing near the energy $\sim1.967$
eV is formed by a bundle of many-body states with significant residues.
The lower branch is instead contributed by one state only at the energy
$\sim1.958$ eV. The energy splitting of the two branches is about
$9$ meV and is less than the Rabi coupling $\Omega=20$ meV. The
small energy splitting is again attributed to the reduced oscillator
strength, which we now turn to discuss in greater detail.

As we mentioned earlier, a plausible picture for the formation of
trion-polaritons is the strong effective light-matter coupling between
a photons and an attractive Fermi polaron of the exciton impurity.
It is clear that only the free part of the attractive polaron (as
characterized by $\phi_{0}$) contribute to the light-matter coupling,
in the form of the term $(\Omega/2)[a_{0}^{\dagger}(\phi_{0}X_{0})+h.c.]$
at zero momentum. In other words, the effective Rabi coupling would
be given by 
\begin{equation}
\Omega_{\textrm{eff}}\simeq\Omega\phi_{0}=\Omega\sqrt{Z_{X}},
\end{equation}
which is reduced by the square root of the excitonic residue. This
expression of the effective Rabi coupling would also work well for
the repulsive polaron (i.e., the exciton-polariton with the electron
gas).

In Fig. \ref{fig:fig4_GroundStatePolaron}, we show the ground-state
energy of the trion-polariton (a) and its excitonic and photonic residues
(b), as a function of the photon detuning $\delta$. The excitonic
residue does not change significant when $\delta\geqslant\varepsilon_{X}$.
In particular, at the avoided crossing of $\delta=\varepsilon_{X}$,
the excitonic residue $Z_{X}\sim0.25$, which implies an effective
Rabi coupling $\Omega_{\textrm{eff}}\simeq\Omega\sqrt{Z_{X}}=10$
meV, which is very close to the observed value of $9$ meV. The slightly
reduced Rabi coupling of $17$ meV at the avoided crossing of the
exciton-polariton might be understand in a similar way. We may identify
that the excitonic residue of the repulsive polaron at $\delta=\varepsilon_{X}$
is about $Z_{X}\sim0.7$. Therefore the effective Rabi coupling is
given by $\Omega_{\textrm{eff}}\simeq\Omega\sqrt{Z_{X}}=16.7$ meV,
in agreement with our finding.

\begin{figure}
\begin{centering}
\includegraphics[width=0.5\textwidth]{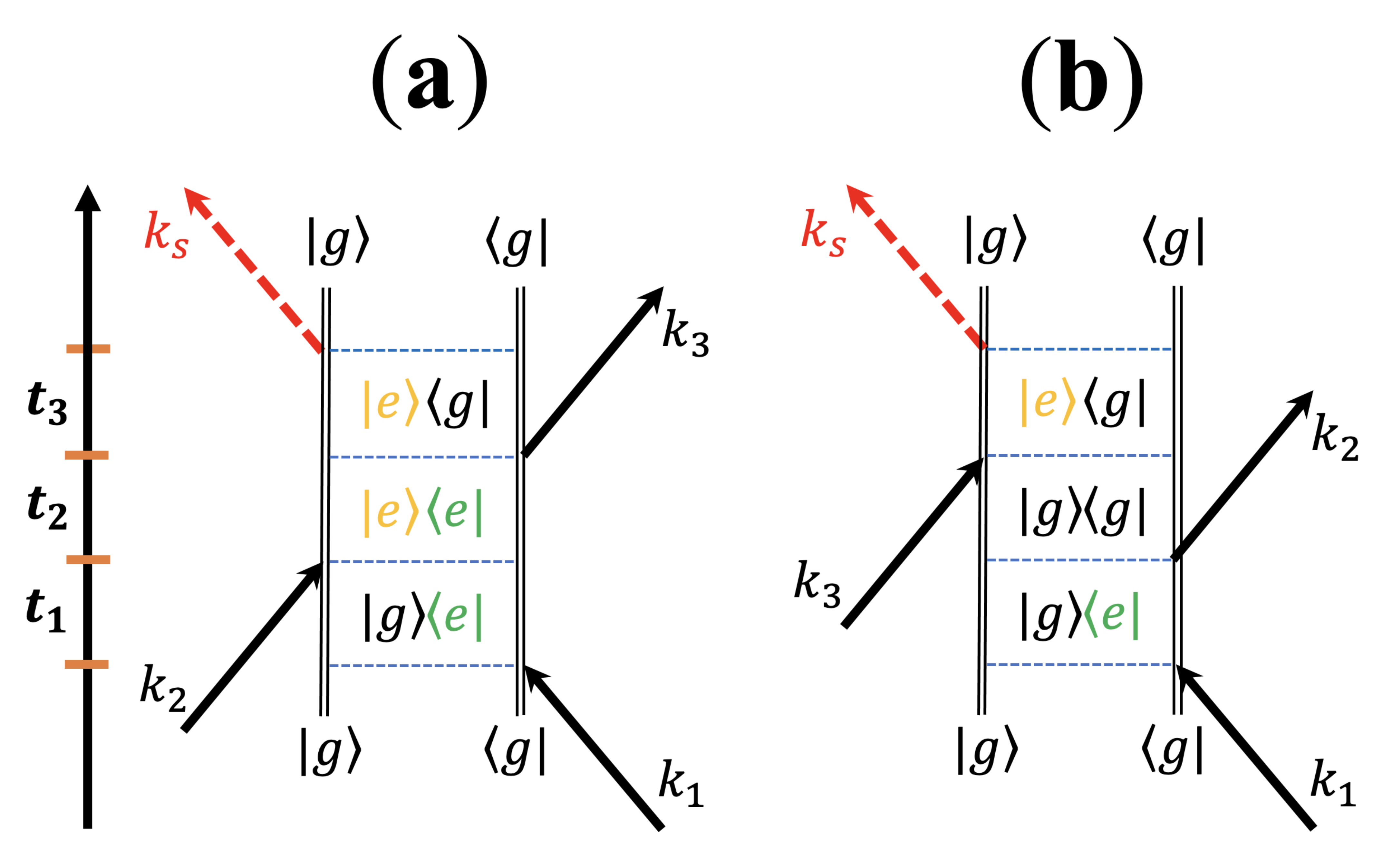}
\par\end{centering}
\centering{}\caption{\label{fig:fig5_2DCS_sketch} Two double-sided Feynman diagrams that
represent the two contributions to the standard rephasing 2D coherent
spectra under the phase-match condition $\mathbf{k}_{s}=-\mathbf{k}_{1}+\mathbf{k}_{2}+\mathbf{k}_{3}$,
with the time ordering of excitation pulses indicated on the left.
The evolution, mixing, and emission time delays are labeled as $t_{1}$,
$t_{2}$, and $t_{3}$, respectively. (a) shows the process of excited-state
emission (ESE), $R_{2}(t_{1},t_{2},t_{3})$. (c) corresponds to the
ground-state bleaching (GSB), $R_{3}(t_{1},t_{2},t_{3})$. In the
diagrams, we use $\left|g\right\rangle $ to denote the Fermi sea
and $\left|e\right\rangle $ to label the many-body states with an
exciton-polariton, respectively. There are infinitely many many-body
(Fermi polaron) states $\left|e\right\rangle $, as indicated by different
colors.}
\end{figure}

\section{two-dimensional coherent spectroscopy}

Let us now consider the 2DCS spectroscopy, which is to be implemented
in future experiments on studying the exciton-polariton physics in
TMD materials. In 2DCS, three excitation pulses with momentum $\mathbf{k}_{1}$,
$\mathbf{k}_{2}$ and $\mathbf{k}_{3}$ are applied to the system
under study at times $\tau_{1}$, $\tau_{2}$ and $\tau_{3}$, separated
by an evolution time delay $t_{1}=\tau{}_{2}-\tau_{1}$ and a mixing
time delay $t_{2}=\tau_{3}-\tau_{2}$, as illustrated in the left
part of Fig. \ref{fig:fig5_2DCS_sketch}. These pulses generate a
signal with momentum $\mathbf{k}_{s}$, as a result of the nonlinear
third-order process of four-wave-mixing. The signal can then be measured
after an emission time delay $t_{3}$ by using the frequency-domain
heterodyne detection. 

During the excitation period, each excitation pulse creates or annihilates
an exciton. As the photon momentum of the excitation pulses is negligible,
the exciton has the zero momentum. Therefore, each pulse can be described
by the interaction operator $V$, 
\begin{equation}
V\propto X_{0}+X_{0}^{\dagger}.
\end{equation}
Following the standard nonlinear response theory \cite{Cho2008},
the four-wave-mixing signal is given by the third-order nonlinear
response function, 
\begin{equation}
\mathcal{R}^{(3)}\propto\left\langle \left[\left[\left[V\left(t_{1}+t_{2}+t_{3}\right),V\left(t_{1}+t_{2}\right)\right],V\left(t_{1}\right)\right],V\right]\right\rangle ,
\end{equation}
where the time-dependent interaction operator $V(t)\equiv e^{i\mathcal{H}t}Ve^{-i\mathcal{H}t}$,
and $\left\langle \cdots\right\rangle $ stands for the quantum average
over the initial many-body configuration of the system without excitation
pulses, which at zero temperature is given by the ground state. By
expanding the three bosonic commutators, we find four distinct correlation
functions and their complex conjugates \cite{Cho2008}. For the rephasing
mode that is of major experimental interest, $t_{1}>0$ and $\mathbf{k}_{s}=-\mathbf{k}_{1}+\mathbf{k}_{2}+\mathbf{k}_{3}$.
For this case, only two contributions are relevant if we consider
at most one excitonic excitation in the system: the process of so-called
excited-state emission (ESE) \cite{Cho2008,Hao2016NanoLett} , 
\begin{equation}
R_{2}=\left\langle VV\left(t_{1}+t_{2}\right)V\left(t_{1}+t_{2}+t_{3}\right)V\left(t_{1}\right)\right\rangle ,
\end{equation}
and the process of ground-state bleaching (GSB) \cite{Cho2008,Hao2016NanoLett},
\begin{equation}
R_{3}=\left\langle VV\left(t_{1}\right)V\left(t_{1}+t_{2}+t_{3}\right)V\left(t_{1}+t_{2}\right)\right\rangle .
\end{equation}
These two processes can be visualized by using double-sided Feynman
diagrams, as given in Fig. \ref{fig:fig5_2DCS_sketch}(a) and Fig.
\ref{fig:fig5_2DCS_sketch}(b), respectively.

For an exciton system, a microscopic calculation of the 2DCS spectrum
has been recently carried out \cite{Hu2022arXiv}. Here, we extend
such a microscopic calculation to the exciton-polariton system. After
some straightforward algebra following the line of Ref. \cite{Hu2022arXiv},
we obtain the ESE and GSB contributions, 
\begin{eqnarray}
R_{2} & = & \sum_{nm}Z_{X}^{(n)}Z_{X}^{(m)}e^{i\mathcal{E}^{(n)}t_{1}}e^{i\left[\mathcal{E}^{(n)}-\mathcal{E}^{(m)}\right]t_{2}}e^{-i\mathcal{E}^{(m)}t_{3}},\\
R_{3} & = & \sum_{nm}Z_{X}^{(n)}Z_{X}^{(m)}e^{i\mathcal{E}^{(n)}t_{1}}e^{-i\mathcal{E}^{(m)}t_{3}},
\end{eqnarray}
where the indices $n$ and $m$ run over the whole many-body polaron
states. 

These two expressions can be easily understood from the double-sided
Feynman diagrams. For the ESE process illustrated in Fig. \ref{fig:fig5_2DCS_sketch}(b),
the weight $Z_{X}^{(n)}Z_{X}^{(m)}$ measures the transfer rates between
different many-body states induced by the three excitation pulses
and the four-wave-mixing signal. For example, the transfer of the
first pulse at momentum $\mathbf{k}_{1}$ brings a factor of $\phi_{0}^{(n)}$,
while the transfer of the second pulse at momentum $\mathbf{k}_{2}$
comes with a factor of $[\phi_{0}^{(m)}]^{*}$, and so on. When we
combine all the four factors for the four transitions, we obtain the
weight $Z_{X}^{(n)}Z_{X}^{(m)}$. On the other hand, the three dynamical
(time-evolution) phase factors arise from the phases accumulated during
the time delays $t_{1}$, $t_{2}$ and $t_{3}$, respectively. The
GSB process can be analyzed in an exactly same way. The only difference
is the absence of the mixing time ($t_{2}$) dependence in the expression.
This is easy to understand from Fig. \ref{fig:fig5_2DCS_sketch}(b):
between the second and third pulses the system returns to the ground
state of a Ferm sea, so there is no phase accumulation during the
mixing time delay.

By taking a double Fourier transformation for $t_{1}$ and $t_{3}$
in $R_{2}(t_{1},t_{2},t_{3})$ and $R_{3}(t_{1},t_{2},t_{3})$, we
obtain the 2DCS spectrum \cite{Hu2022arXiv},
\begin{equation}
\mathcal{S}\left(\omega_{1},t_{2},\omega_{3}\right)=\sum_{nm}\frac{Z_{X}^{(n)}Z_{X}^{(m)}}{\left(-\omega_{1}\right)^{-}-\mathcal{E}^{(n)}}\frac{1+e^{i\left[\mathcal{E}^{(n)}-\mathcal{E}^{(m)}\right]t_{2}}}{\omega_{3}^{+}-\mathcal{E}^{(m)}},\label{eq:3rdResponseS}
\end{equation}
where $(-\omega_{1})^{-}\equiv-\omega_{1}-i0^{+}$, and $\omega_{1}$
and $\omega_{3}$ are the excitation energy and emission energy, respectively.

\begin{figure*}
\begin{centering}
\includegraphics[width=0.3\textwidth]{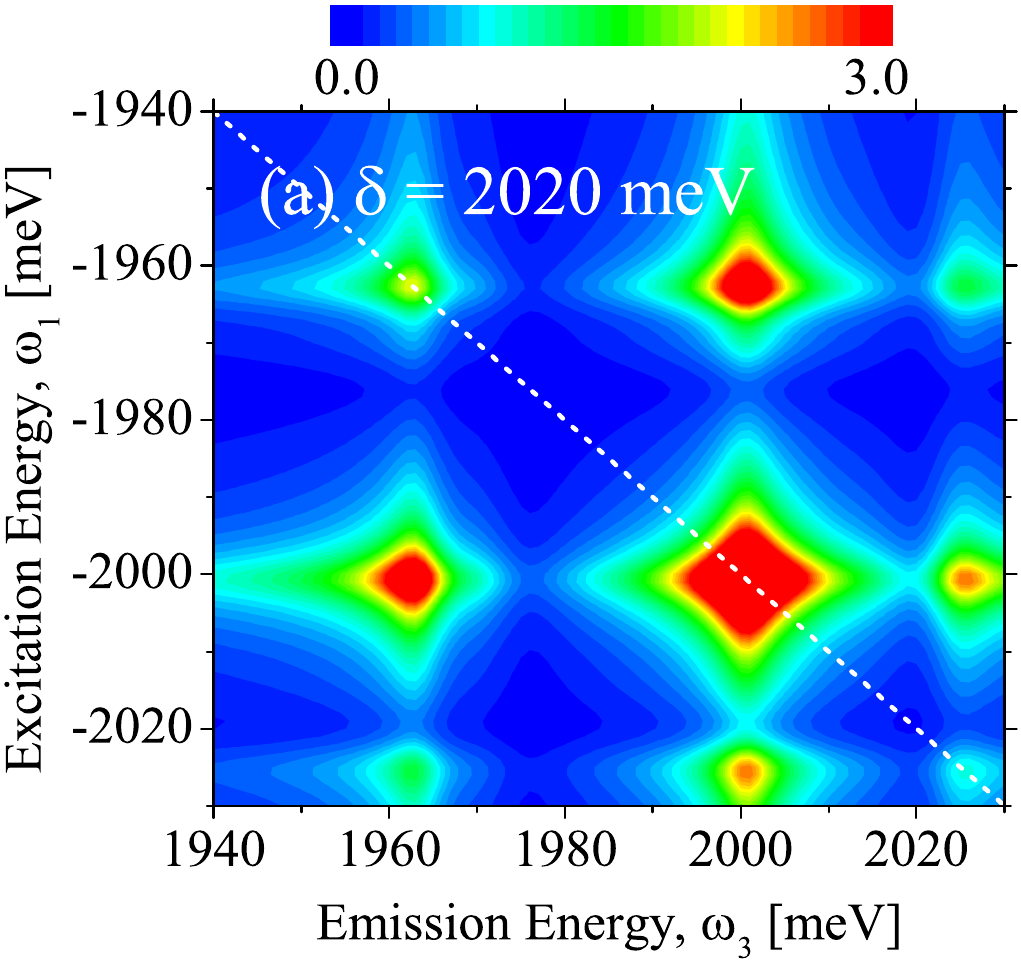}\includegraphics[width=0.3\textwidth]{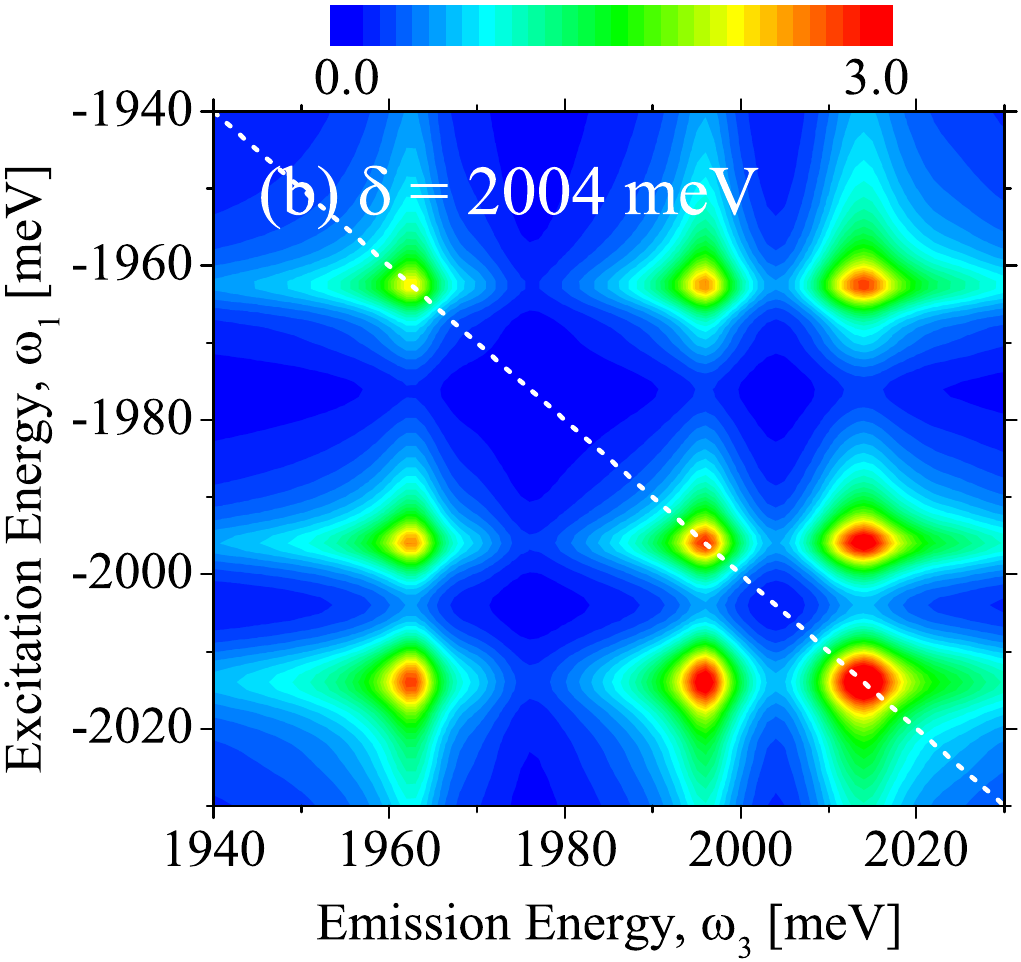}\includegraphics[width=0.3\textwidth]{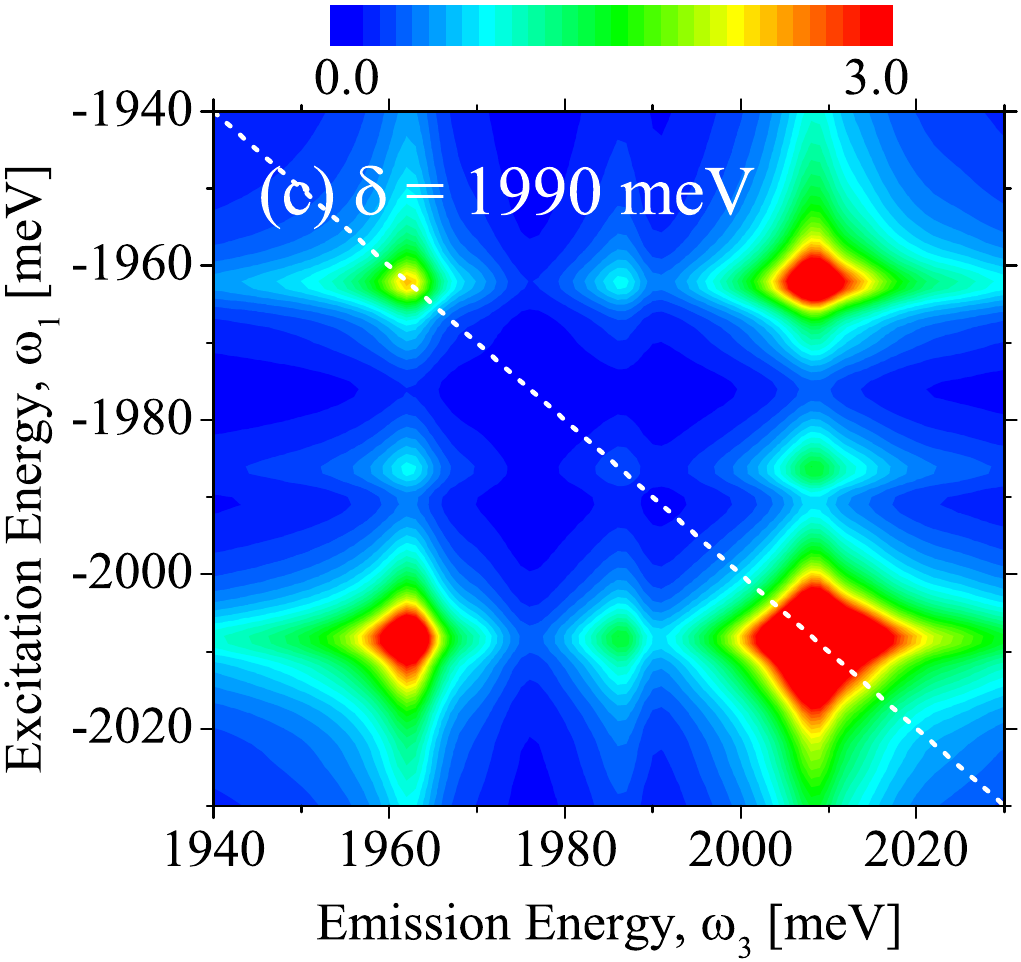}
\par\end{centering}
\begin{centering}
\includegraphics[width=0.3\textwidth]{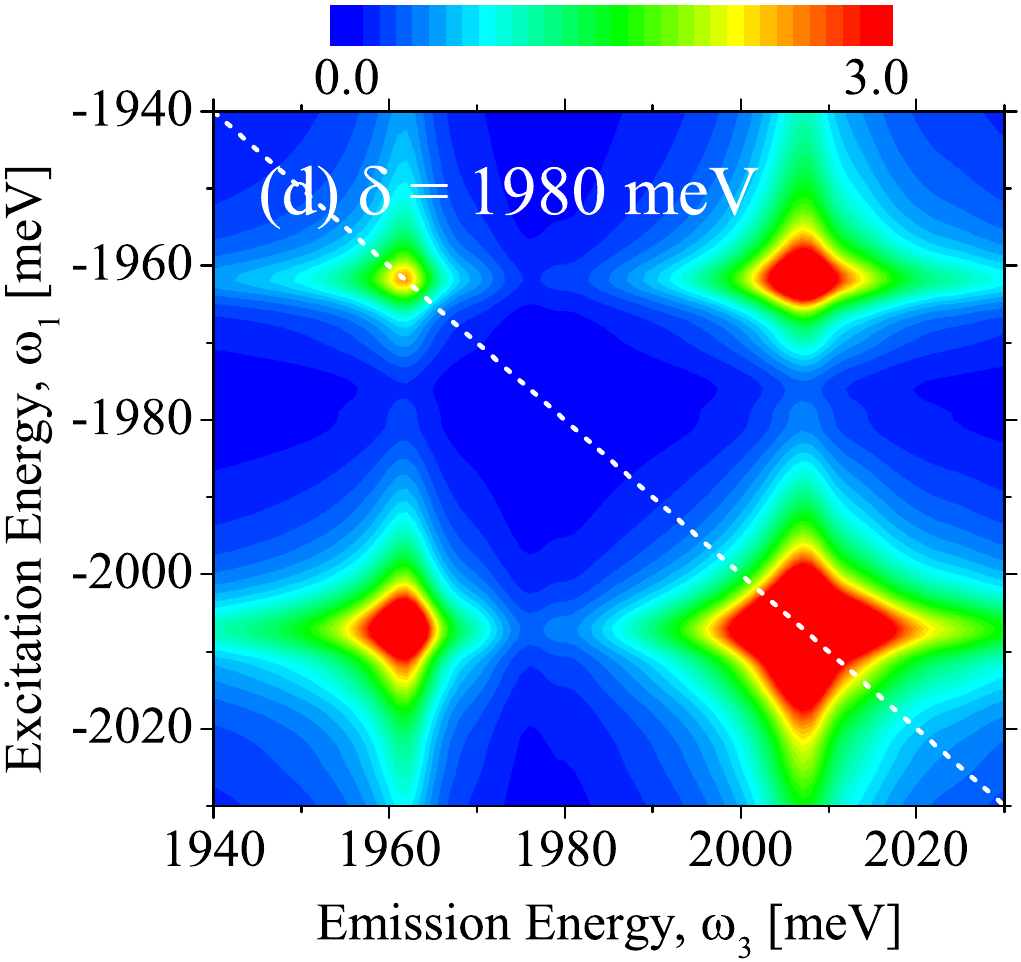}\includegraphics[width=0.3\textwidth]{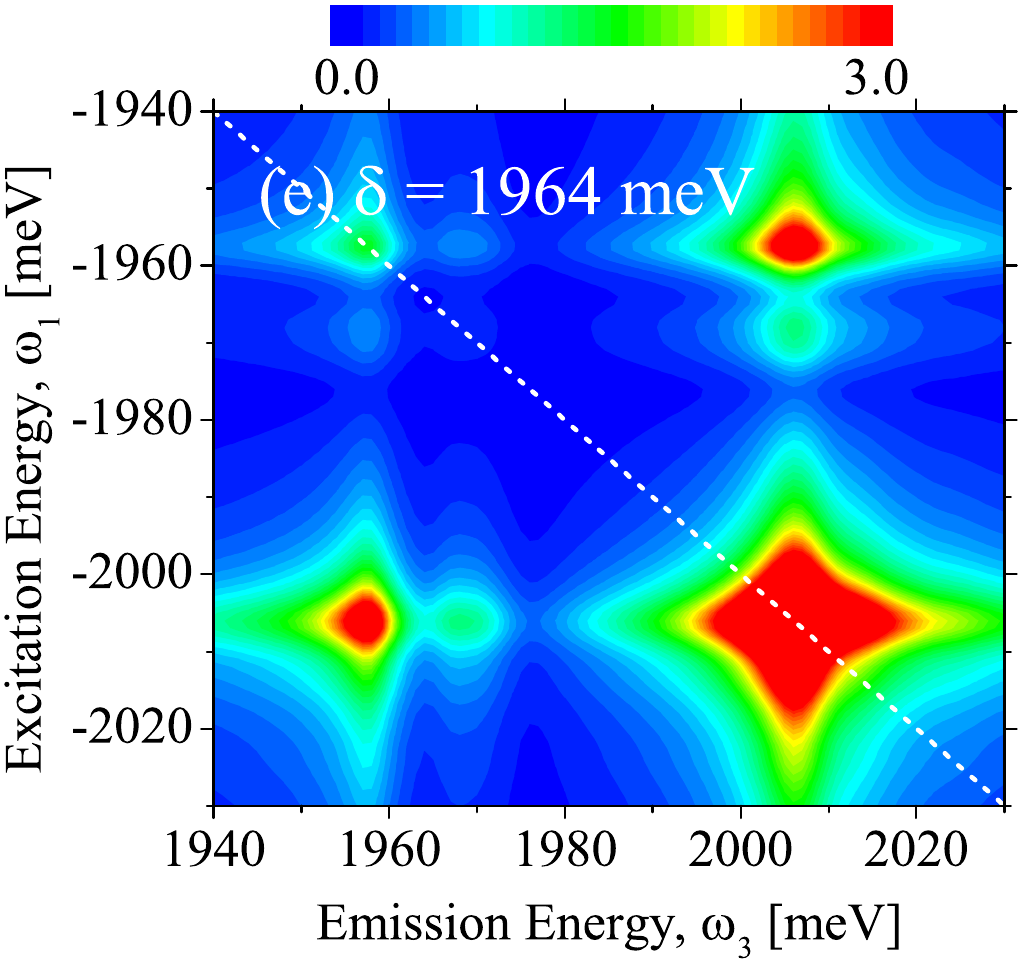}\includegraphics[width=0.3\textwidth]{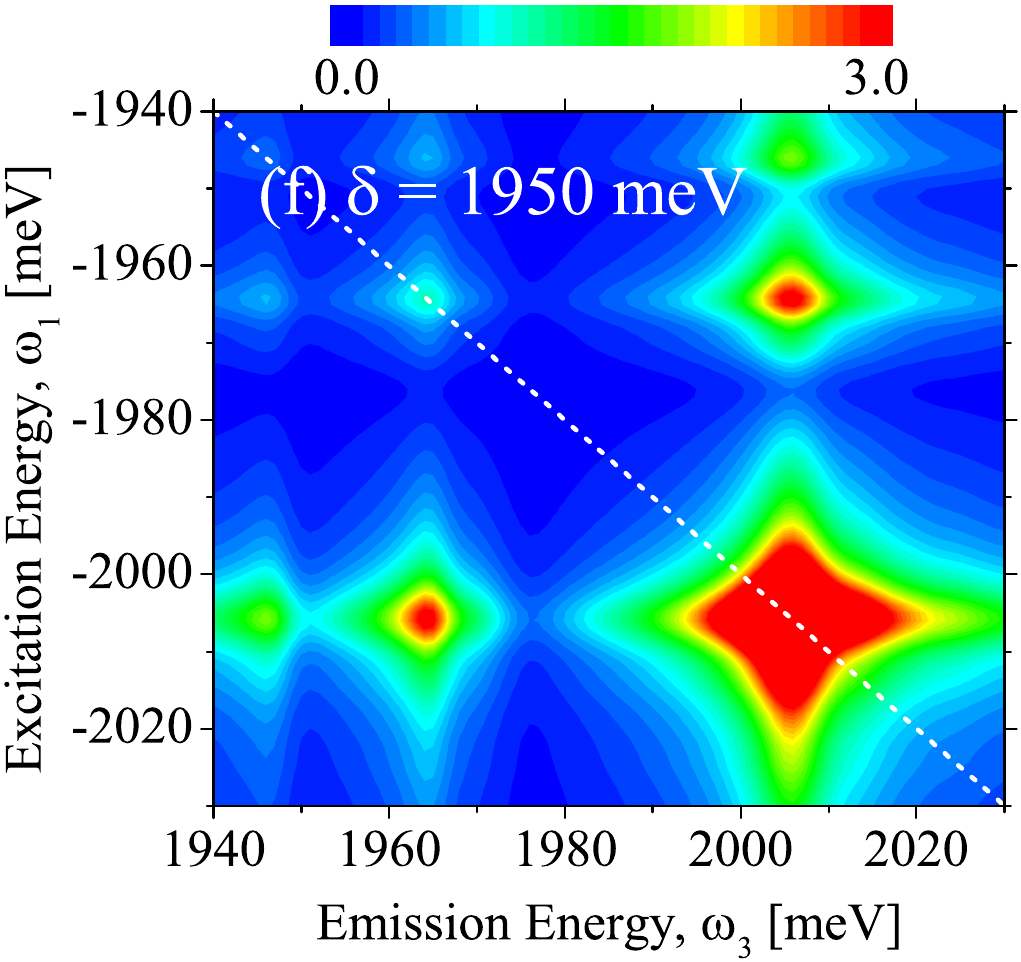}
\par\end{centering}
\centering{}\caption{\label{fig:fig6_2DCS0} The simulated rephasing 2D coherent spectra
(amplitude) at various photon detunings and at zero mixing time decays
$t_{2}=0$. The photon detuning decreases from $\delta=2020$ meV
to from $\delta=1950$ meV in (a)-(f). We typically find three peaks
appearing on the diagonal dashed line. The red color illustrates the
maximum amplitude, as indicated in the colormap above each subplot.
The electron Fermi energy is set to be $\varepsilon_{F}=7.8$ meV.}
\end{figure*}

\subsection{Zero mixing time delay $t_{2}=0$}

Let us first focus on the case of zero mixing time delay $t_{2}=0$,
where 
\begin{equation}
\mathcal{S}\left(\omega_{1},0,\omega_{3}\right)=2\sum_{nm}\frac{Z_{X}^{(n)}}{\left(-\omega_{1}\right)^{-}-\mathcal{E}^{(n)}}\frac{Z_{X}^{(m)}}{\omega_{3}^{+}-\mathcal{E}^{(m)}},\label{eq:S0t2}
\end{equation}
and consider the dependence of the 2DCS spectrum $\left|\mathcal{S}(\omega_{1},0,\omega_{3})\right|$
on the photon detuning $\delta$, as shown in Fig. \ref{fig:fig6_2DCS0}.
By changing $\delta$ from the blue shift above the exciton-polariton
crossing (a) to the red shift below the trion-polariton crossing (f),
we typically find three diagonal peaks located at the diagonal line
$\omega_{3}=-\omega_{1}$ (see the white dashed lines) and six off-diagonal
cross-peaks located symmetrically with respect to the diagonal line.

These peaks arise from the three branches of excitations, as we already
seen in Fig. \ref{fig:fig1_spectrum}. Formally, with decreasing energy
the many Fermi-polaron-polariton states have been grouped into the
upper polariton, middle polariton, and lower polariton branches, as
often referred to in the literature \cite{Sidler2017,Tan2020,Rana2021,BastarracheaMagnani2021,Zhumagulov2022}.
Therefore, we can roughly understand the Fermi-polaron-polariton as
a three-energy-level system, with the energies $\mathcal{E}^{(n)},\mathcal{E}^{(m)}\sim E_{UP}$,
$E_{MP}$, and $E_{LP}$ that are tunable by the cavity photon detuning.
The corresponding \emph{excitonic} weights are given by the excitonic
residues $Z_{X}^{(UP)}$, $Z_{X}^{(MP)}$, and $Z_{X}^{(LP)}$. Hence,
from Eq. (\ref{eq:S0t2}) we can easily identify that the diagonal
peaks occur when $\omega_{1}=-E_{\alpha}$ and $\omega_{3}=E_{\alpha}$
with peak amplitude $(Z_{X}^{(\alpha)})^{2}$ ($\alpha=UP,MP,LP$),
while the off-diagonal peaks appear when $\omega_{1}=-E_{\alpha}$
and $\omega_{3}=E_{\beta}$ with peak amplitude $Z_{X}^{(\alpha)}Z_{X}^{(\beta)}$
($\alpha\neq\beta=UP,MP,LP$).

The experimental measurement of diagonal peaks and crossover peaks
at zero mixing time delay $t_{2}=0$ then provides us the information
of both the energies $E_{\alpha}$ and the residues $Z_{X}^{(\alpha)}$.
In particular, when the photon detuning $\delta$ is near the two
avoided crossings (as shown in Fig. \ref{fig:fig6_2DCS0}(b) and \ref{fig:fig6_2DCS0}(e),
respectively), we may easily identify the effective Rabi coupling
from the corresponding energy splitting. Near the avoided crossing
for the trion-polariton, the asymmetry of the crossing can also be
clearly seen. 

We note that, when the photon detuning is tuned to roughly the half-way
between the two avoided crossing, the middle polariton disappears
in the 2DCS spectrum (see Fig. \ref{fig:fig6_2DCS0}(d)). This is
simply because, at this detuning the middle polariton is of photonic
in characteristics, with negligible excitonic component. Therefore,
it can not be seen from the 2DCS, which probes the excitonic part
instead of the photonic part of the system. This feature of the 2DCS
spectrum is useful to characterize the main component of the ground-state
of the trion-polariton (or the lower polariton branch). As the photon
detuning decreases across the avoided crossing for trion-polaritons,
we find that the brightness of the diagonal trion-polariton peak becomes
much weaker.

\begin{figure*}
\begin{centering}
\includegraphics[width=0.3\textwidth]{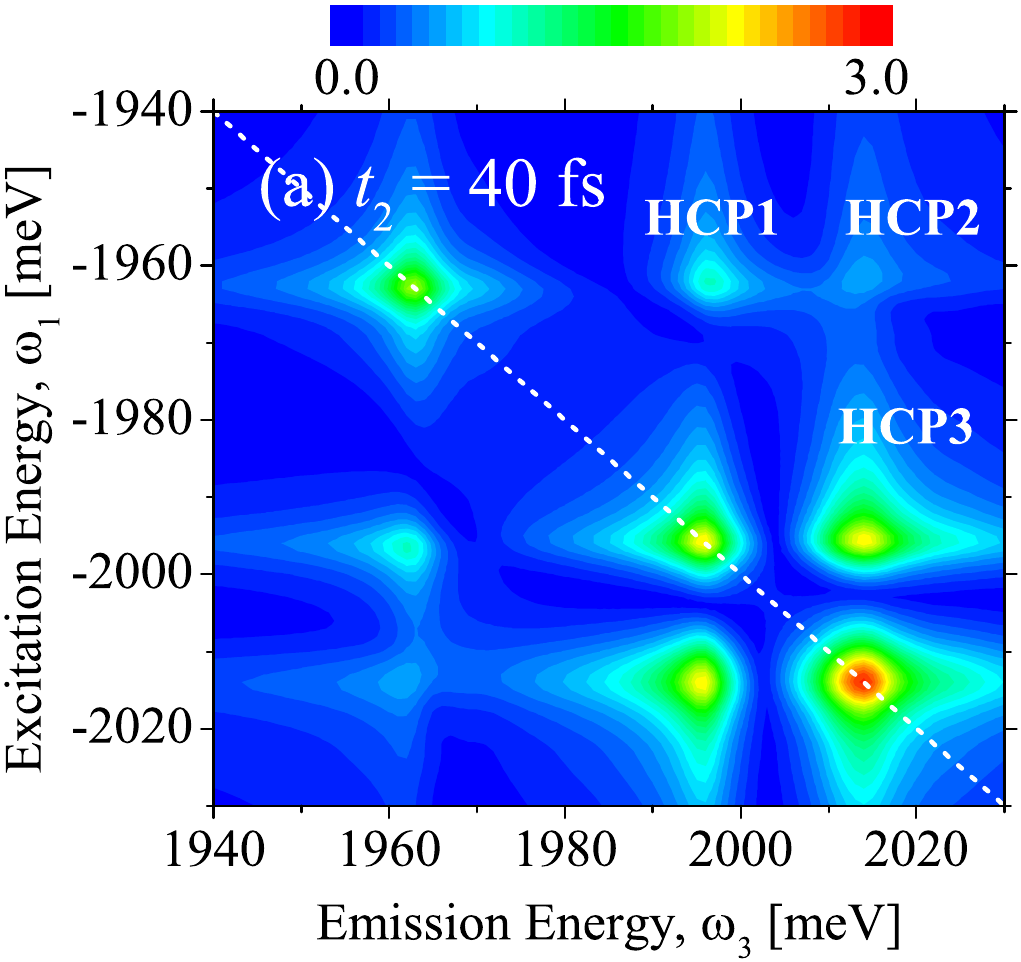}\includegraphics[width=0.3\textwidth]{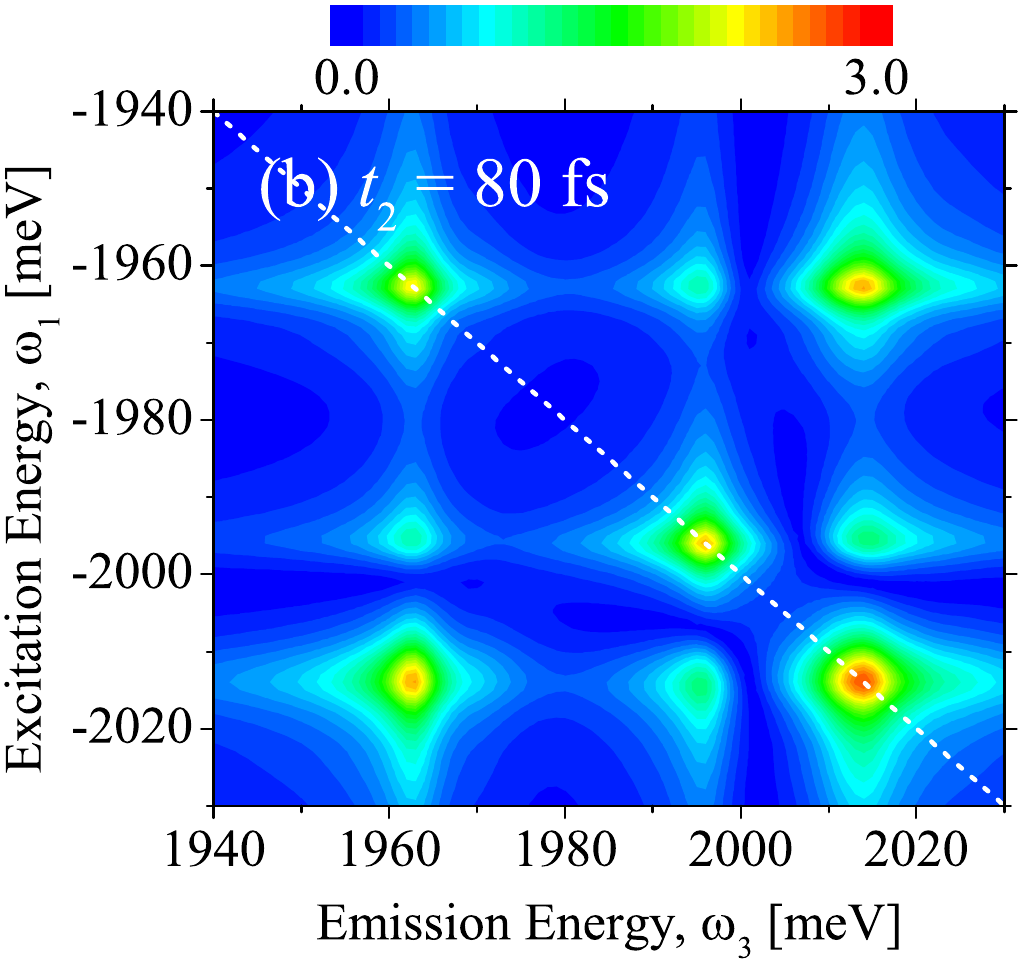}\includegraphics[width=0.3\textwidth]{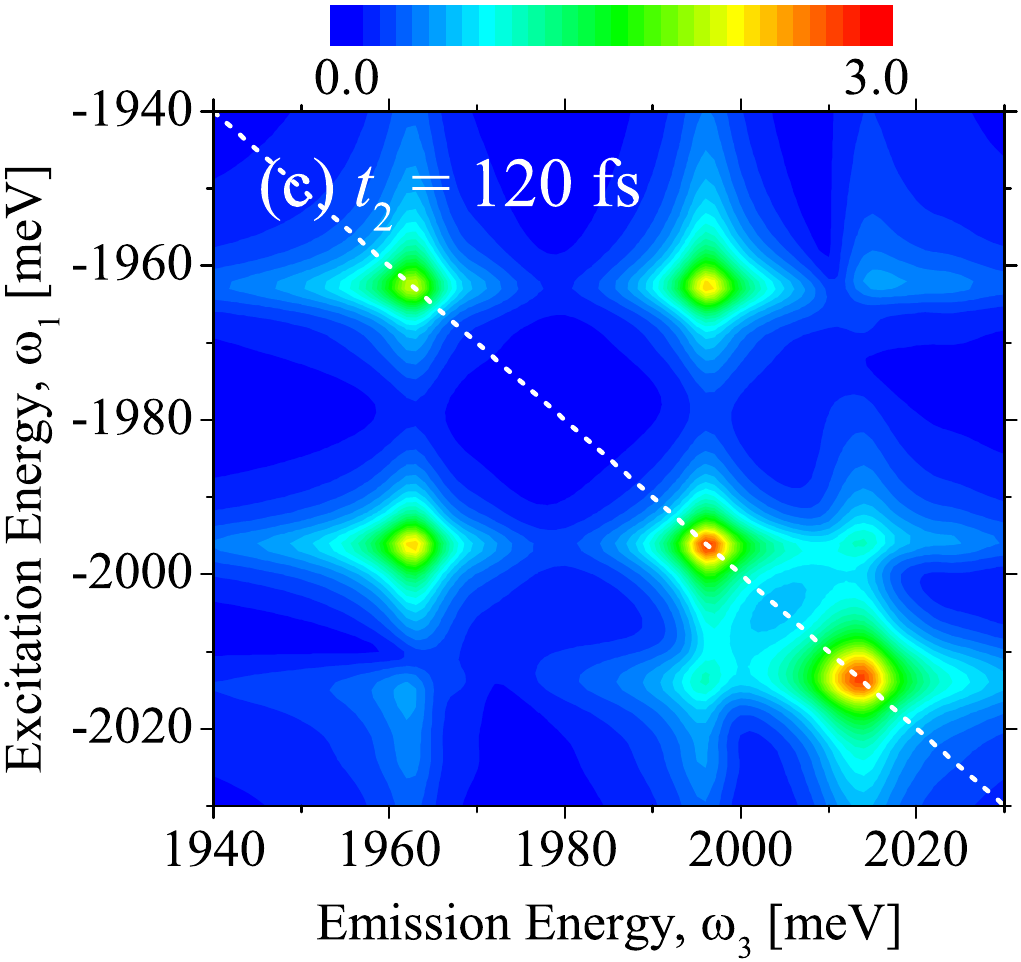}t
\par\end{centering}
\begin{centering}
\includegraphics[width=0.3\textwidth]{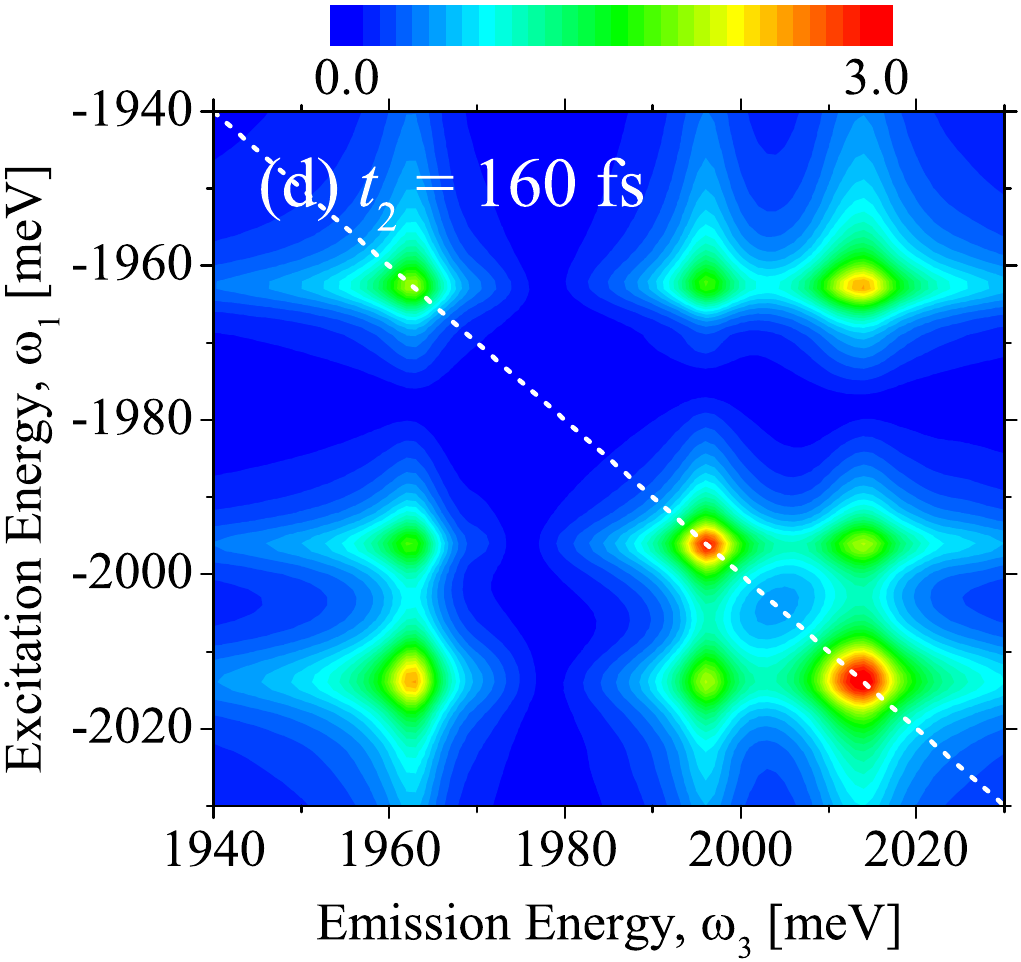}\includegraphics[width=0.3\textwidth]{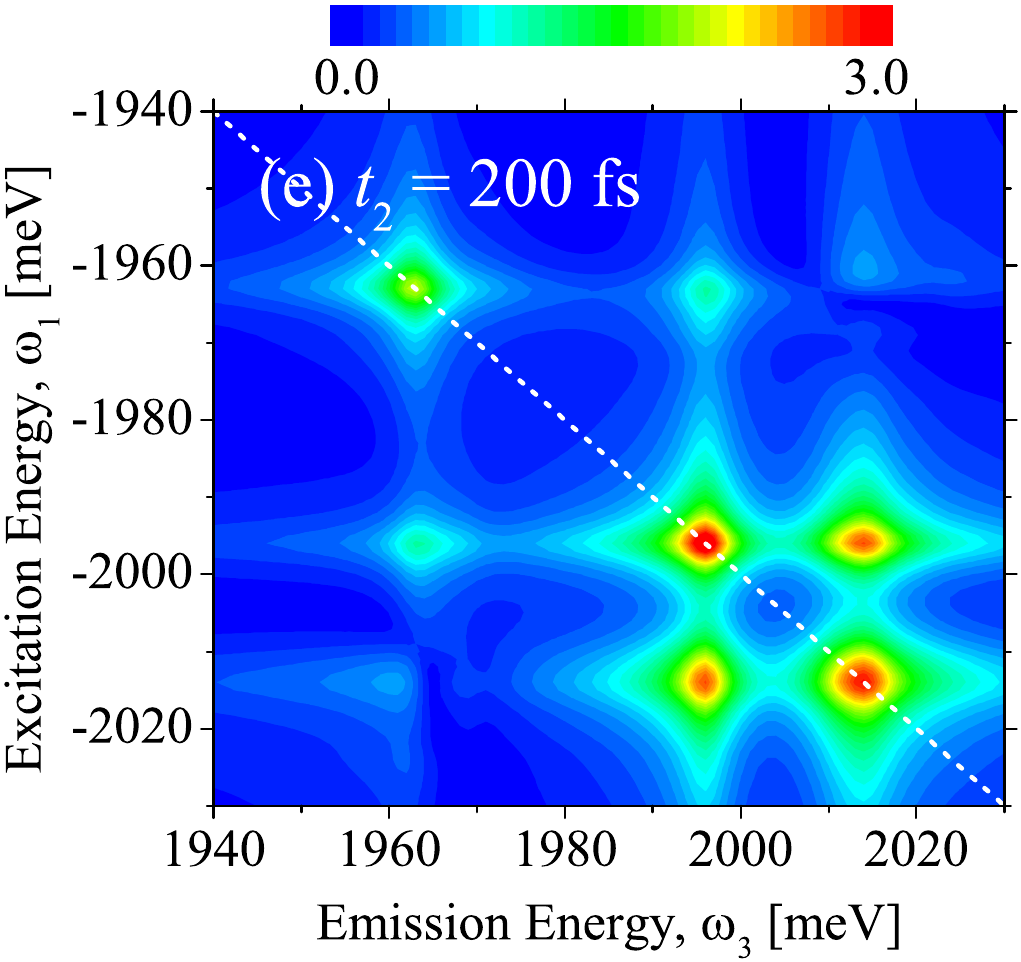}\includegraphics[width=0.3\textwidth]{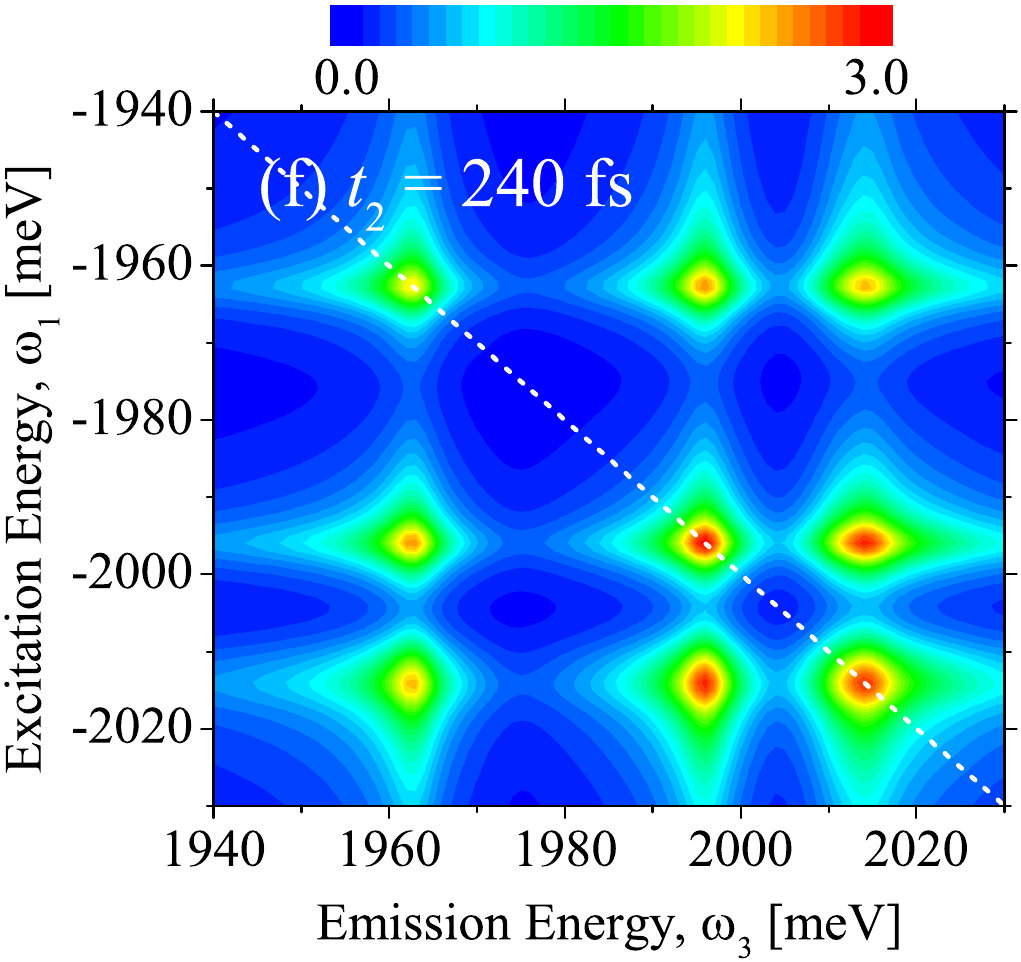}
\par\end{centering}
\centering{}\caption{\label{fig:fig7_2DCSt2} The simulated rephasing 2D coherent spectra
(amplitude) at the photon detuning $\delta=2004$ meV with increasing
mixing time decays $t_{2}$ from (a) to (f). In (a), the three higher-cross-peak
(HCP) are indicated. The three peaks appearing on the diagonal dashed
line essentially do not change. However, the higher-cross-peaks and
lower-cross peaks oscillate as a function of $t_{2}$, revealing the
quantum coherence among different quasiparticles. The red color illustrates
the maximum amplitude, as indicated in the colormap above each subplot.
The electron Fermi energy is set to be $\varepsilon_{F}=7.8$ meV.}
\end{figure*}

Although the upper, middle and lower polariton branches can also be
conveniently measured by using one-dimensional optical response, such
as the reflectance spectroscopy and photoluminescence spectroscopy
\cite{Sidler2017}, the application of 2DCS spectroscopy has unique
features to discriminate the intrinsic homogeneous line-with of the
resonance peaks \cite{Hao2016NanoLett} and the interaction effects
\cite{Li2006,Muir2022}. Unfortunately, both effects (i.e., the disorder
potential for excitons and the exciton-exciton interaction) are not
included in our model Hamiltonian. Nevertheless, our results in Fig.
\ref{fig:fig6_2DCS0} provide the essential qualitative features of
the 2DCS spectrum, which is to be measured in future exciton-polariton
experiments. In addition, the appearance of the off-diagonal peaks
and their evolution as a function of the mixing time decay $t_{2}$
are useful to characterize the quantum coherences among the different
branches of polaritons, which we now turn to discuss.

\begin{figure*}
\centering{}\includegraphics[width=0.3\textwidth]{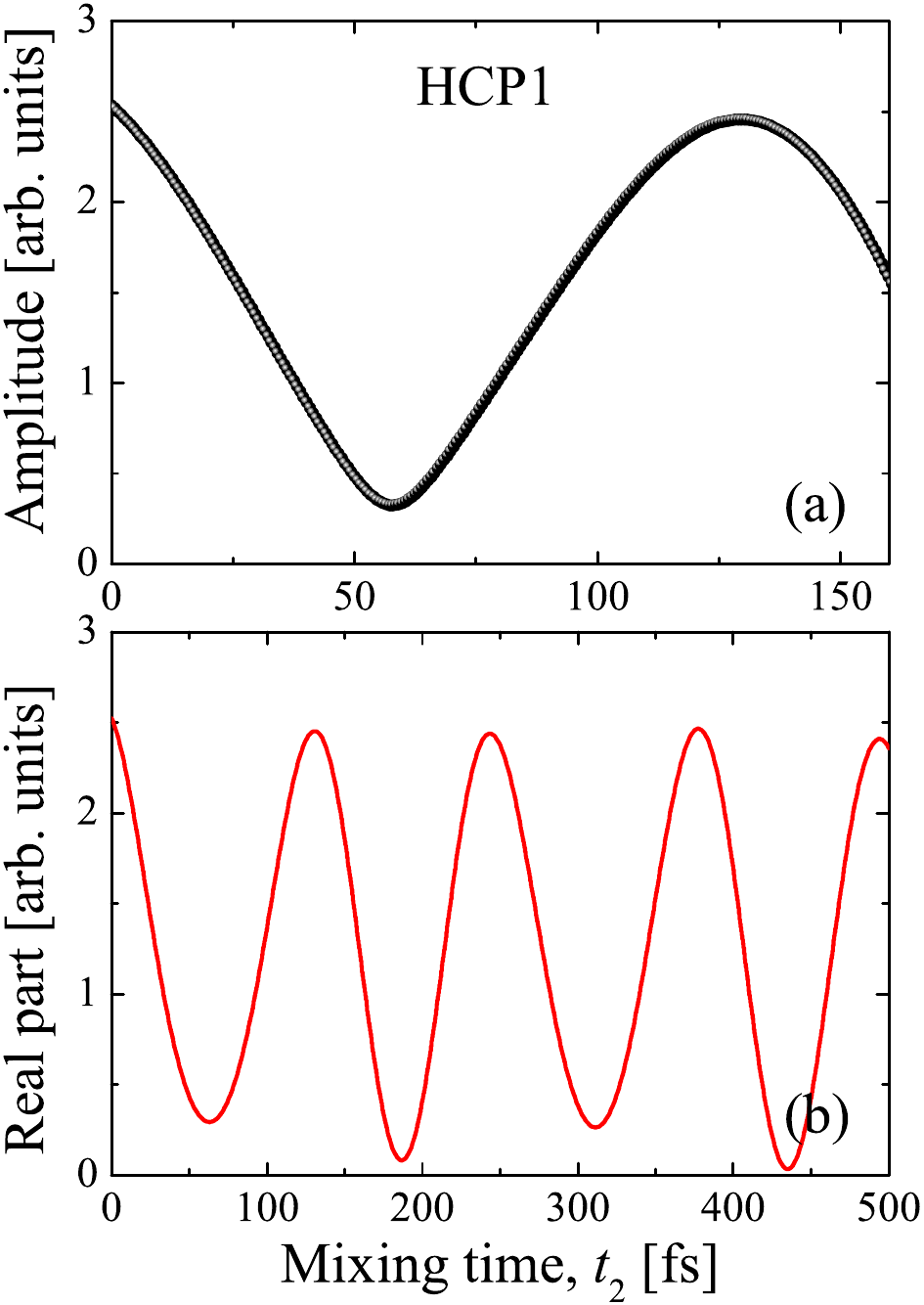}\includegraphics[width=0.3\textwidth]{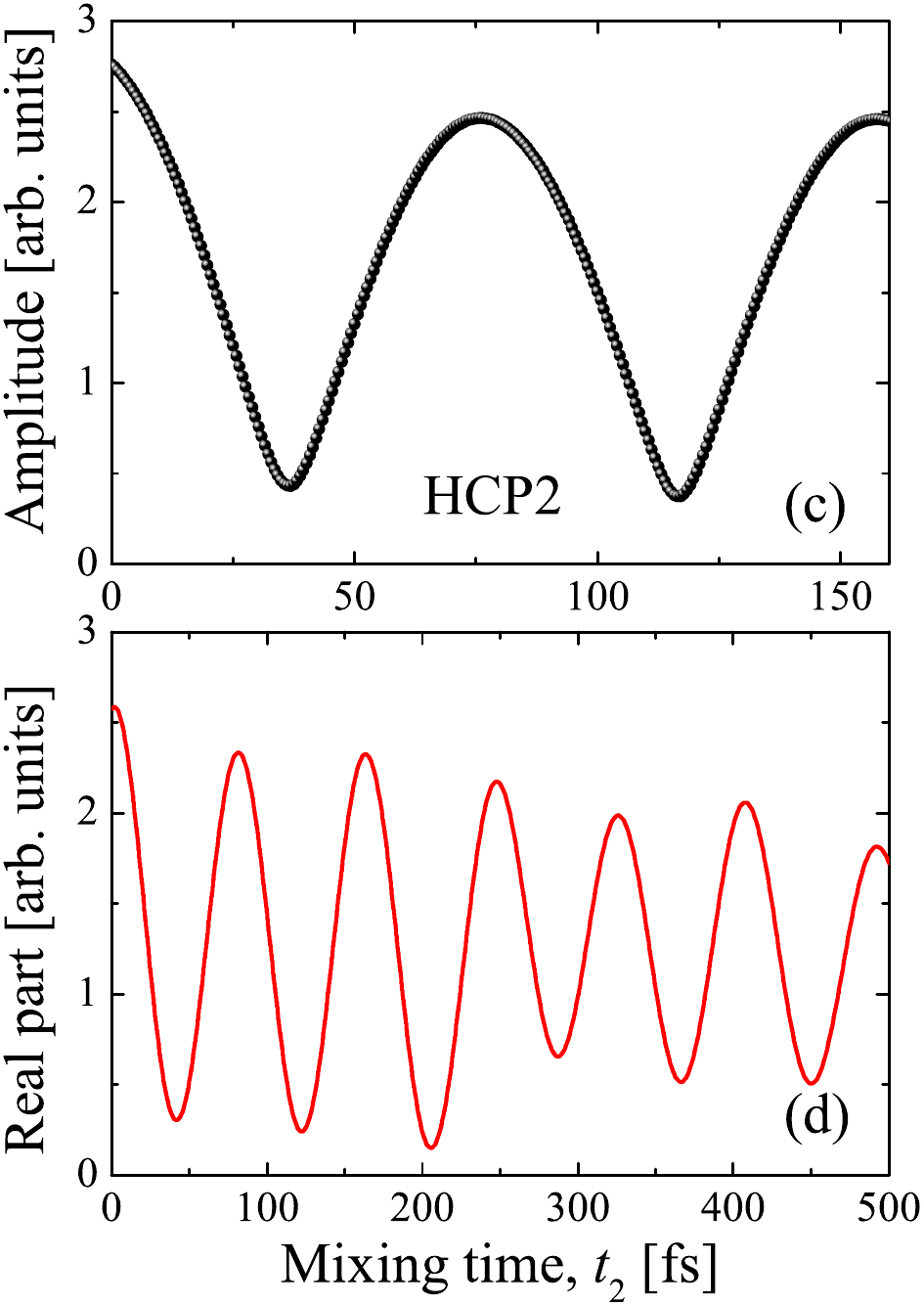}\includegraphics[width=0.3\textwidth]{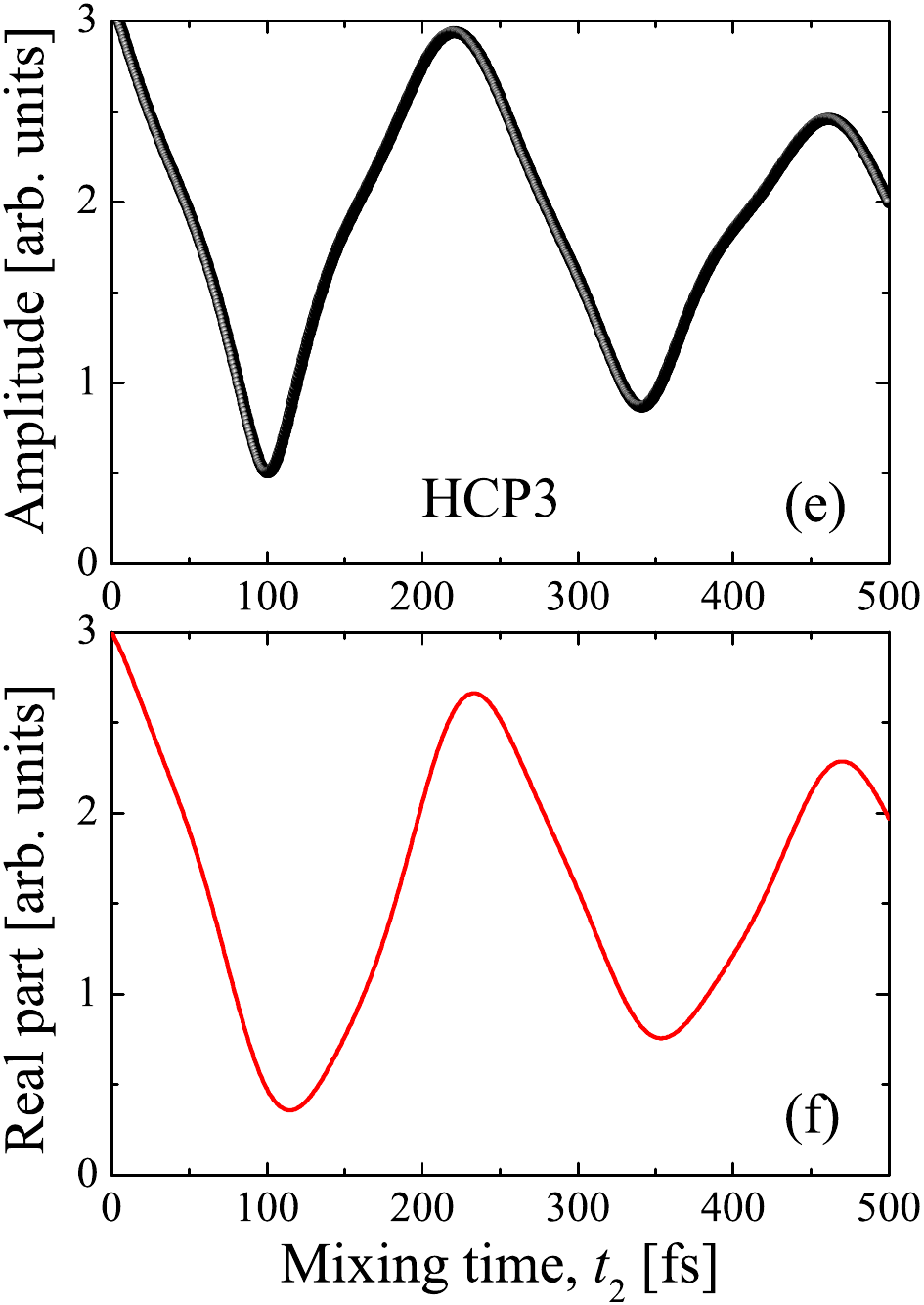}\caption{\label{fig:fig8_crosspeak} The simulated amplitude (a) and real part
(b) of the rephasing 2D signal at the three cross-peaks as a function
of the mixing time delays $t_{2}$. Note the different periodicity
at different crosspeaks. We choose the photon detuning $\delta=2004$
meV as in Fig. \ref{fig:fig6_2DCS0} and take the electron Fermi energy
$\varepsilon_{F}=7.8$ meV.}
\end{figure*}

\subsection{Quantum coherence of the cross-peaks}

In Fig. \ref{fig:fig7_2DCSt2}, we present the simulated rephasing
2D coherent spectra $\left|\mathcal{S}(\omega_{1},t_{2},\omega_{3})\right|$
with increasing mixing time decays $t_{2}$. We choose a photon detuning
$\delta=2004$ meV at the avoided crossing for exciton-polaritons,
where all the three polariton branches are clearly visible at $t_{2}=0$.
We label the three off-peaks at the top-right corner of the figure
as HCP1, HCP2 and HCP3 \cite{Hao2016NanoLett}, respectively.

As can be seen from Eq. (\ref{eq:3rdResponseS}), the time $t_{2}$-dependence
of the 2D spectrum $\mathcal{S}\left(\omega_{1},t_{2},\omega_{3}\right)$
comes in through the term $e^{i\left[\mathcal{E}^{(n)}-\mathcal{E}^{(m)}\right]t_{2}}$.
As we interpret the Fermi-polaron-polariton as a three-level system,
where the energy levels $\mathcal{E}^{(n)},\mathcal{E}^{(m)}$ are
to be replaced by $E_{UP}$, $E_{MP}$, and $E_{LP}$, it is readily
seen that the $t_{2}$-term gives rise to quantum oscillations with
three different periods: $2\pi/\left|E_{MP}-E_{LP}\right|$ for the
HCP1 cross-peak, $2\pi/\left|E_{UP}-E_{LP}\right|$ for HCP2, and
$2\pi/\left|E_{UP}-E_{MP}\right|$ for HCP3. At the photon detuning
$\delta=2004$ meV, we find that $E_{MP}-E_{LP}\simeq33.2$ meV, $E_{UP}-E_{LP}\simeq50.5$
meV, and $E_{UP}-E_{MP}\simeq17.3$ meV. Therefore, the periodicities
of the cross-peaks are at the order of $10^{-10}$ s or $100$ fs,
and are given by $T_{\textrm{HCP1}}\simeq124.4$ fs, $T_{\textrm{HCP2}}\simeq81.6$
fs, and $T_{\textrm{HCP3}}\simeq239.5$ fs.

The 2DCS spectra in Fig. \ref{fig:fig7_2DCSt2} are shown in $40$
fs increments. We can clearly identify that the brightness of each
cross-peak \emph{oscillates} with the mixing time delay $t_{2}$,
revealing the coherent coupling among different branches of exciton-polariton
and trion-polaritons. In comparison with the zero mixing time delay
2DCS in Fig. \ref{fig:fig6_2DCS0}(b), we find that the HPC1 cross-peak
nearly recovers its full brightness at $t_{2}=120$ fs and $240$
fs, confirming that its periodicity is close to the anticipated value
$T_{\textrm{HCP1}}\simeq124.4$ fs. For the HCP2 cross-peak, we see
similarly that it nearly disappears at $t_{2}=40$ fs, $120$ fs and
$200$ fs and fully recovers at t $t_{2}=80$ fs, $160$ fs and $240$
fs, in agreement with our anticipation that $T_{\textrm{HCP2}}\simeq81.6$
fs. In addition, the HPC3 cross-peak only returns to the its full
brightness at $T_{\textrm{HCP3}}\simeq240$ fs.

To better characterize the quantum oscillations, we report in Fig.
\ref{fig:fig8_crosspeak} the simulated rephasing 2D signal at the
crosspeaks as a function of the mixing time $t_{2}$, both in the
form of its amplitude (the upper panel) and in its real part (the
lower panel). The oscillations do not take the exact form of $1+\cos(\omega t_{2})$,
as one may naively anticipate from Eq. (\ref{eq:3rdResponseS}). This
is partly due to the existence and competition of three different
periods in the oscillations, which may bring a slight irregular structure.
On the other hand, we find that the oscillations at HCP2 and HCP3
typically exhibit a decay. These dampings should be related to the
many-body nature of the upper-polariton branch, i.e., it is formed
by a bundle of many-body states as we discussed in Fig. \ref{fig:fig2_ResidueExcitonCrossing}.
Therefore, the upper polariton has an intrinsic spectral broadening,
which eventually causes the damping in the quantum oscillation of
the cross-peaks HCP2 and HCP3. In contrast, both the lower-polariton
and middle-polariton at the detuning $\delta=2004$ meV are dominated
by a single Fermi polaron state, and do not experience the intrinsic
spectral broadening. As a result, the quantum oscillation at HCP1
is long-lived, if we do not take into account the lifetimes of excitons
(due to the natural radiative decay) and of photons (due to the quality
of the cavity).

\section{Conclusions and outlooks}

In conclusions, based on the Fermi polaron description of an exciton-polariton
immersed in an electron gas, we have analyzed the structure of exciton-polaritons
and trion-polaritons in monolayer transition metal dichalcogenides
and have predicted their 2D coherent spectroscopy for on-going experimental
explorations in the near future. 

From the structure analysis, we have found that the upper-polariton
branch at the exciton-polariton avoided crossing typically consists
of a number of many-body Fermi polaron states. Instead, the lower-polariton
branch at the trion-polariton avoided crossing involves only one Fermi
polaron state. The situation for the middle-polariton branch varies,
depending on whether it is close to the exciton-polariton crossing
or close to the trion-polariton crossing. In the former case, the
middle-polariton is also dominated by a single Fermi polaron state.

As there are three polariton branches \cite{Sidler2017,Rana2021,Zhumagulov2022},
in the 2D coherent spectroscopy we have found three diagonal peaks
and six off-diagonal cross-peaks. From these peaks measured in future
experiments, in principle we should be able to extract the excitonic
residues of different polariton branches. We have predicted the existence
of quantum oscillations in the 2D spectra as a function of the mixing
time delay $t_{2}$, as the evidence for the quantum coherence among
the different polariton branches \cite{Hao2016NanoLett}.

Although in the present study we have not considered the effects of
the disorder potential on excitons and the inter-exciton interaction,
our results would provide a good starting point to understand the
2D coherent spectroscopy on exciton-polaritons to be experimentally
measured in the near future. Theoretically, the inclusions of the
disorder effect and interaction effect would be extremely challenging
in numerics, since the dimension of the Hilbert space of the model
Hamiltonian will increase dramatically. We will address these effects
in future publications.

\section{Statements and Declarations}

\subsubsection{Ethics approval and consent to participate }

Not Applicable.

\subsubsection{Consent for publication }

Not Applicable.

\subsubsection{Availability of data and materials }

The data generated during the current study are available from the
contributing author upon reasonable request. 

\subsubsection{Competing interests}

The authors have no competing interests to declare that are relevant
to the content of this article. 

\subsubsection{Funding}

This research was supported by the Australian Research Council's (ARC)
Discovery Program, Grants No. DE180100592 and No. DP190100815 (J.W.),
and Grant No. DP180102018 (X.-J.L).

\subsubsection{Authors' contributions }

All the authors equally contributed to all aspects of the manuscript.
All the authors read and approved the final manuscript. 

\subsubsection{Acknowledgements }

See funding support.

\subsubsection{Authors' information}

Hui Hu, Centre for Quantum Technology Theory, Swinburne University
of Technology, Melbourne 3122, Australia, Email: hhu@swin.edu.au

Jia Wang, Centre for Quantum Technology Theory, Swinburne University
of Technology, Melbourne 3122, Australia, Email: jiawang@swin.edu.au

Riley Lalor, Centre for Quantum Technology Theory, Swinburne University
of Technology, Melbourne 3122, Australia, Email:103192654@student.swin.edu.au

Xia-Ji Liu, Centre for Quantum Technology Theory, Swinburne University
of Technology, Melbourne 3122, Australia, Email: xiajiliu@swin.edu.au

\appendix
\begin{figure*}
\begin{centering}
\includegraphics[width=0.8\textwidth]{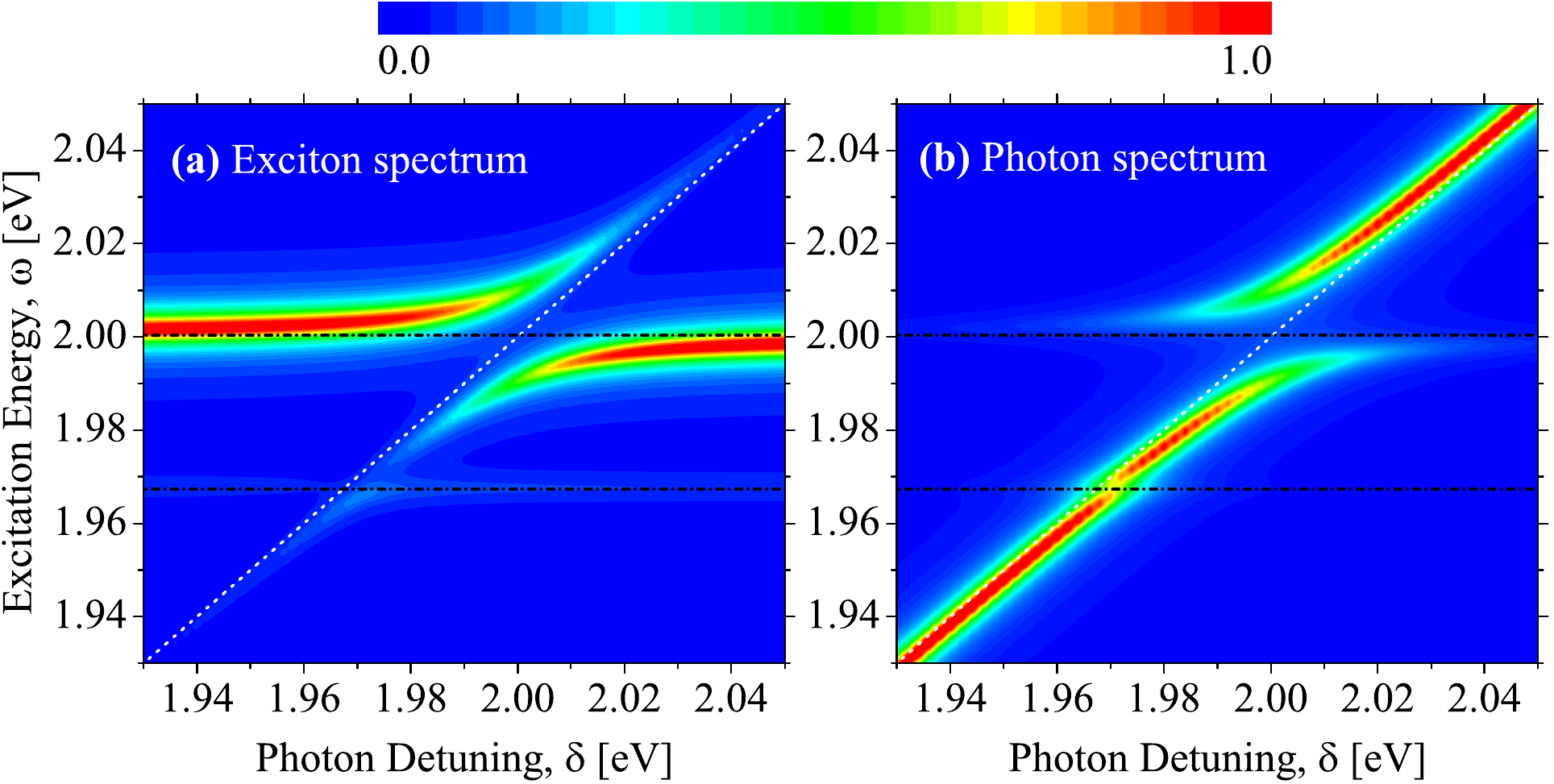}
\par\end{centering}
\caption{\label{fig:figS1_spectrum} Two-dimensional contour plots of the zero-momentum
spectral functions of the exciton (a) and of the photon (b), as a
function of the photon detuning $\delta$ at a small electron Fermi
energy $\varepsilon_{F}=1.0$ meV. The two black horizontal dot-dashed
lines show the energies of the exciton (i.e., the repulsive polaron
branch with $\varepsilon_{X}\simeq2000.4$ meV) and the trion (i.e.,
the attractive polaron branch with $\varepsilon_{T}\simeq1967.4$
meV), in the absence of the cavity photon field. The diagonal white
dotted line indicates the photon detuning $\omega=\delta$. At this
low electron density, the avoided crossing at $\omega=E_{T}$ is insignificant.
The spectral functions are measured in arbitrary units and are plotted
in a linear scale..}
\end{figure*}

\section{Exciton-polaritons and trion-polaritons at small electron density}

In Fig. \ref{fig:figS1_spectrum}, we report the zero-momentum spectral
functions of excitons and of photons for a small electron density
with Fermi energy $\varepsilon_{F}=1.0$ meV. At this density, the
exciton energy level is barely affected by the scattering with the
electron gas and the trion energy level is basically given by the
trion binding energy of $E_{T}\simeq32$ meV. We can hardly identify
the existence of the trion-polariton from the excitonic spectrum.
Neither, the trion-polariton can barely be seen from the photonic
spectrum. Both spectra are very similar to the spectrum of exciton-polaritons
in the \emph{absence} of the electron gas.

\end{document}